\documentclass[%
 reprint,
superscriptaddress,
nofootinbib,
amsmath,amssymb,
aps,
]{revtex4-1}
\pdfoutput=1

\usepackage[pdftex]{graphicx}
\usepackage{amssymb}
\usepackage{enumitem}
\usepackage[linktocpage,breaklinks]{hyperref}
\usepackage[dvipsnames]{xcolor}
\usepackage{etoolbox}

\usepackage{hyperref}
\usepackage{orcidlink}

\hypersetup{colorlinks=true,
            citecolor=RoyalBlue,
            linkcolor=RoyalBlue,
            urlcolor=RoyalBlue}

\usepackage{orcidlink}
\usepackage{xspace}
\usepackage{array}
\usepackage{dcolumn}
\newcolumntype{d}[1]{D{.}{.}{#1}}
\usepackage{longtable}

\makeatletter
\newcommand{\mainmatter}{%
  \setcounter{footnote}{0}%
  \patchcmd{\@makefntext}{\fnsymbol}{\arabic}{}{}%
  \patchcmd{\@thefnmark}{\fnsymbol}{\arabic}{}{}%
  \def\@makefnmark{\textsuperscript{\arabic{footnote}}}%
}
\makeatother

\graphicspath{{./figs/}}

\newcommand{\orcid}[1]{\href{https://orcid.org/#1}{\includegraphics[height=\fontcharht\font`\B]{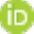}}}

\newcommand{\fuka}{\texttt{FUKA}\xspace}

\newcommand{\kadath}{\texttt{KADATH}\xspace}

\newcommand{\eg}{e.g.,\xspace}
\newcommand{\D}{\mathrm{d}}
\newcommand{\hrns}{\texttt{Hydro-RNS}\xspace}
\newcommand{\rns}{\texttt{RNS}\xspace}
\newcommand{\etk}{\texttt{Einstein Toolkit}\xspace}
\newcommand{\uteen}{\texttt{U13}\xspace}
\newcommand{\sbsix}{\texttt{SB6}\xspace}
\newcommand{\dduteen}{\texttt{SLy.U13}\xspace}
\newcommand{\ddsbsix}{\texttt{SLy.SB6}\xspace}
\newcommand{\grhayl}{\texttt{GRHayL}\xspace}
\newcommand{\grhayleos}{\texttt{GRHayLEOS}\xspace}
\newcommand{\igm}{\texttt{IllinoisGRMHD}\xspace}
\let\l=\left
\let\r=\right

\newcommand{\gammaij}{\gamma_{ij}}
\newcommand{\confgammaij}{\tilde{\gamma}_{ij}}

\newcommand{\tD}{\tilde{D}}

\newcommand{\eqrefs}[2]{Eq.\,(\eqref{#1} -- \eqref{#2})}

\definecolor{newpurple}{HTML}{951CED}

\newcommand{\uidaho}{Department of Physics, University of Idaho, Moscow, ID 83843, USA}
\newcommand{\goetheuni}{Institute for Theoretical Physics, Goethe University, Frankfurt, Germany}
\newcommand{\wvu}{Department of Physics and Astronomy, West Virginia University, Morgantown, WV 26506, USA}
\newcommand{\cgwc}{Center for Gravitational Waves and Cosmology, West Virginia University, Chestnut Ridge Research Building, Morgantown, WV 26505}

\begin{document}
\sloppy 

\date{\today}

\title[S. Tootle et al]{A new code for computing differentially rotating neutron stars}
\author{Samuel D. Tootle}
\affiliation{\uidaho}
\affiliation{\goetheuni}
\author{Terrence Pierre Jacques}
\affiliation{\uidaho}
\affiliation{\wvu}
\affiliation{\cgwc}
\author{Marie Cassing}
\affiliation{\goetheuni}

\begin{abstract}
    We present new initial data codes for constructing stationary,
    axisymmetric equilibrium models of differentially rotating neutron
    stars in full general relativity within the Frankfurt
    University/KADATH (\fuka) suite of initial data codes. \fuka
    leverages the \kadath spectral library to solve the Einstein
    equations under the assumption of an isentropic fluid without
    magnetic fields while incorporating \grhayleos to support 3D
    tabulated equations of state in \textit{stellar collapse} format. The
    two solvers explored in this work include one using quasi-isotropic
    coordinates (QIC) in Spherical coordinates while the other solves the
    eXtended Conformal Thin Sandwich (XCTS) decomposition in Cartesian
    coordinates, enabling the construction of equilibrium configurations
    with high accuracy and efficiency. In this work we adopt the
    Komatsu-Eriguchi-Hachisu differential rotation law, however, the code
    is designed to be extensible to other rotation laws, allowing for
    exploration of physically relevant sequences and critical rotation
    thresholds. Furthermore, we perform convergence tests demonstrating
    the exponential accuracy of the spectral approach, we validate QIC
    and XCTS solutions against models well-studied in the literature, and
    we also compare \fuka solutions against the well-known \rns code.
    Finally, we explore the impact that initial data resolution has on
    dynamical simulations and recover the convergence order of the
    evolution scheme, the dominate source of error in this study. The new
    \fuka codes and results presented here lay the foundation for future
    extensions to more general configurations, including magnetic fields,
    removal of isentropic assumptions, and binary systems, and have been
    made publicly available to support community efforts in modeling
    differentially rotating relativistic stars.
\end{abstract}
\maketitle

\section{Introduction}
\label{sec:intro}
Gravitational-wave astronomy, particularly when combined with
electromagnetic observations, has transformed our ability to probe the
dynamics and microphysics of compact stellar objects~\cite{Abbott2017b,
LIGOScientific2017vwq,LIGOScientific2018cki,LIGOScientific2018hze}. Many
astrophysical scenarios, such as core-collapse supernovae
\cite{Ott06c,Burrows07a}, accretion-induced collapse~\cite{Dessart2006,
Abdikamalov2009aq}, and binary neutron-star (BNS) mergers
\cite{Shibata2003ga, Baiotti08, Baiotti2016}, can lead to the formation
of hot, rapidly and differentially rotating
remnants~\cite{Hanauske2016,Reboul-Salze:2024jst, Bamber:2024qzi,
Chawhan:2025esv, Rainho:2025ykl}. These remnants can encounter
non-axisymmetric instabilities, motivating the critical need for accurate
equilibrium models and diagnostics that connect quasi-equilibrium
sequences to dynamical behavior~\cite{Rezzolla00, Shibata:2002mr,
Saijo2003,Cassing2024}. However, developing accurate models for these
systems requires high-accuracy numerical relativity (NR) simulations
capable of capturing dynamics that span several length and time scales.
The first step towards enabling accurate and efficient studies of systems
in these unique configurations is the calculation of initial data
solutions that satisfy the general relativistic constraint equations, the
focus of this work.

A computationally less costly approach to studying rotating stars is to
assume uniform rotation. Previous studies of rotating equilibria have
shown the existence of (i) a Keplerian stability line, beyond which mass
shedding takes place, and (ii) the existence of a turning-point stability
line.  While (ii) is an indication of an unstable solution, it is not
necessarily a sufficient condition for the onset of a dynamical
instability that will eventually lead to gravitational
collapse~\cite{Friedman88, Takami2014}. More importantly, the maximum
mass allowed by uniform rotation and the maximum mass of the
corresponding non-rotating configuration show a proportionality and even
universal relation~\cite{Cook92b, Lasota1996, Breu2016, Musolino2023b}.

Constructing stationary, axisymmetric equilibria with differential
rotation requires closing the relativistic Euler integral with a rotation
law. The Komatsu–Eriguchi–Hachisu (KEH) prescription~\cite{Komatsu89b}-
often termed the $j$-constant law — has long served as the baseline for
modeling differential rotation because of its computational simplicity
and the existence of an analytic solution to the first integral of
motion.
In addition to providing valuable insight into the impact of differential
rotation on stellar structure and stability, the KEH model serves as a
test bed for new initial data codes to validate their implementation.
This includes a wide array of public initial data codes such as
\rns~\cite{Stergioulas95}, \texttt{XNS}\footnote{Support for cold
tabulated EOS has become available in \texttt{XNS}v4}~\cite{Pili2017,
Franceschetti2022}, and \texttt{LORENE}~\cite{lorene} as well as private
initial data codes such as \texttt{COCAL}~\cite{Uryu2017}. Public initial
data codes have played an essential role in enabling systematic studies
of differentially rotating sequences, stellar equilibrium properties
~\cite{Goussard1998, Baumgarte00b, Ansorg2009, Gondek2016,
Studzinska2016, Bozzola2017, Weih2017, Zhou2019, Espino2019, Bozzola2019,
Zhou2019, Franceschetti2022, Staykov2023, Cheong2024, Nishad2024} and
their dynamical behavior~\cite{Duez:2006qe, Giacomazzo2011, Weih2017,
Espino2019b}.
At the same time, sustained progress has broadened the rotation-law
landscape. Motivated by profiles extracted from merger simulations, where
the angular velocity often peaks off-axis, generalized prescriptions that
specify $\Omega(j)$ rather than $j(\Omega)$ have been introduced and
explored in detail, including the multi-parameter families proposed by
Uryu+~\cite{Uryu2017} and subsequent variants~\cite[see
\eg][]{Passamonti2020, Xie2020, Camelio2021, Cassing2024}. Such laws
allow for better control over core and envelope regions and facilitate
targeted comparisons to merger-inspired equilibria~\cite{Kuan2024}.

Contributions to the public initial data code landscape are essential to
validate the accuracy of previous results, provide additional support for
models with predictive power, and enable the broader community to
participate in novel exploration. In this spirit, this work aims to
contribute to this rich community and to chart a path towards areas of
advancements to broaden our ability to explore astrophysical systems of
interest involving differentially rotating neutron stars, using the
Frankfurt University/KADATH (\fuka) initial-data framework.

\fuka is a suite of multi-purpose, multi-physics initial data (ID)
solvers that leverage spectral methods to solve the Einstein constraint
equations for a variety of astrophysical scenarios. \fuka is built on the
\kadath spectral library~\cite{Grandclement2009} and has been utilized to
compute initial data for highly mass-asymmetric and spinning binary black
holes (BBH), binary neutron stars (BNS), and black hole-neutron star
binaries~\cite{Papenfort2021b}. \fuka has been instrumental in reliably
exploring extreme regions of the space of parameters for BNS and BHNS
configurations, enabled by its robust solving framework, high accuracy,
and support for piecewise polytropic and cold tabulated equations of
state (EOS). An exhaustive list of publications enabled by \fuka is
routinely updated here~\cite{fukascience}.
\fuka also provides a robust, minimal interface for exporting ID
solutions to evolution codes, allowing easy access to the interpolated
solution. To date, \fuka has been coupled to the latest state of the art
codes including, the \etk~\cite{EinsteinToolkit:2025_05},
\texttt{SpECTRE}~\cite{spectrecode}, \texttt{BAM}~\cite{Markin2023a},
\texttt{SACRA}~\cite{Kuan2023a},
\texttt{SPHINCS\_BSSN}~\cite{Rosswog2024a},
\texttt{MHDuet}~\cite{Palenzuela2025}, and
\texttt{GRoovy}~\cite{Jacques2024} as well as others~\cite{Corman2024}.

In this work we take a first step towards providing high-accuracy ID for
differentially rotating neutron stars by extending the \fuka suite of ID
solvers~\cite{Papenfort2021b} to compute isolated neutron star (NS) ID
with differential rotation, while maintaining the high accuracy and
efficiency of spectral methods. To do so, we leverage Quasi-Isotropic
Coordinates (QIC) and the eXtended Conformal Thin Sandwich (XCTS)
formulations to solve the Einstein constraint equations. These approaches
help to minimize the computational resources required for ID generation
and allow for multiple means of self-validation. To this end, we have
extended \fuka to compute isolated NS solutions using QIC, and extended
both QIC and XCTS NS solvers to support differential rotation via the KEH
law. Here QIC is particularly useful as it reduces the system of
equations to essentially a 2D problem, but is restricted to axisymmetric
solutions, whereas XCTS is a generic formulation that can be applied to
non-axisymmetric configurations.

In this work, we aim to: (i) compute and validate high-order, spectrally
accurate KEH solutions using the QIC and XCTS formulations within \fuka;
(ii) establish code-verification pathways against published results and
independent tools that have historically adopted KEH-based sequences; and
(iii) validate the accuracy and effectiveness of each implementation
using dynamical simulations. To do so, we build upon {\kadath}'s
multi-domain spectral scheme with compactified outer boundaries, and on
{\fuka}'s robust solving framework. Furthermore, we extend {\fuka}'s
existing EOS support to include the use of 3D tabulated EOS using
\texttt{GRHayLEOS}~\cite{Cupp2025}, the first known implementation in a
public code.

The remainder of this paper is organized as follows: Section
\ref{sec:math} reviews the mathematical framework (3+1 decomposition, QIC
and XCTS formulations, hydro-stationary equilibrium and the KEH rotation
law). Section \ref{sec:impl} describes the numerical implementation of
the initial data solvers. Section \ref{sec:methods} outlines the methods
used to conduct dynamical simulations using the computed ID solutions.
Section \ref{sec:results} reports verification and validation results,
including cross-formulation analysis and comparison with published
results. Finally, section \ref{sec:conclusion} summarizes our findings
and outlines possible avenues for future research and extensions to
\fuka.

\section{Mathematical Background}
\label{sec:math}
To simulate astrophysical events using general relativity a key mechanism
is to decompose Einstein's field equations using the standard $3+1$
decomposition whereby the spacetime line element takes the
form~\cite{Gourgoulhon2007,Baumgarte2010}
\begin{align}
  {ds^2} &= -\alpha^2 \D t^2
    + \gamma_{ij}(\D x^i + \beta^i \D t)(\D x^j + \beta^j \D t) \,, \label{equ:3dline} \\
  {n_\mu} &\equiv - \alpha \nabla_\mu t \,, \\
  {\gamma_{\mu \nu}} &= g_{\mu \nu} + n_\mu n_\nu \,, \label{equ:3dmetric}
\end{align}
where $\alpha$ is known as the \textit{lapse}, $\beta^i$ is the
\textit{shift}, $n_\mu$ is the unit normal to the three dimensional hypersurface $\Sigma_t$,
and $\gamma_{ij}$ is the purely spatial metric compatible
with $\Sigma_t$.
Using this decomposition, Einstein's field equations can then be recast
into the form
\begin{align}
  {R + K^2 - K_{ij}K^{ij}} &= 16 \pi E \,, \label{equ:Hconst}\\
  {D_j K^j_{\hspace{2mm}i} - D_i K} &= 8 \pi P_i \,, \label{equ:Mconst}
\end{align}
which is an elliptic system of equations regarded as the initial value
problem in numerical relativity. Here $R$ is the Ricci scalar, $K_{ij}$
is the extrinsic curvature of $\Sigma_t$, and $K$ is the trace of
$K_{ij}$, commonly referred to as the mean curvature. The right hand side
(RHS) includes source terms of the total energy $E$ and the momentum
current $P_i$, which are defined using projections of the energy-momentum
tensor ${T^{\mu \nu}}$.

We approximate stellar material as a perfect fluid, where ${T^{\mu \nu}}$
takes the form
\begin{equation}
  {T^{\mu \nu}} \equiv \rho h u^\mu u^\nu + p g^{\mu \nu} \,,
\end{equation}
with $\rho$ being the rest-mass density, $h \equiv 1 + \epsilon + p /
\rho$ the relativistic specific enthalpy, $u^\mu$ the fluid
four-velocity, and $p$ the pressure. By decomposing $T^{\mu \nu}$ using
the $3+1$ decomposition, we obtain the following contractions along
time-like and space-like directions
\begin{subequations}
  \label{equ:SYS_T}
  \begin{align}
    {E} &\equiv T_{\mu\nu}n^\mu n^\nu \,, \label{equ:Eproj} \\
    {P_\alpha} &\equiv -T_{\mu\nu}\gamma^\mu _{\hspace{2mm}\alpha} n^\nu \,, \label{equ:currentproj} \\
    {S_{\alpha\beta}} &\equiv T_{\mu\nu}\gamma^\mu _{\hspace{2mm}\alpha} \gamma^\nu _{\hspace{2mm}\beta} \,. \label{equ:Sproj}
  \end{align}
  \end{subequations}
With these projections and a choice of coordinates, we can endeavour to
compute the initial data solution for an astrophysical system of interest
at a given instant in time.

To compute a stationary, axisymmetric spacetime, it is necessary to
define a Killing vector field $\xi^\mu$ that exhibits these symmetries.
This is equivalent to finding a vector field that is a solution to
Killing's equation
\begin{align}
  {\nabla_{\xi} g_{\mu\nu} }= 0 \,.
\end{align}
Under a 3+1 decomposition, we can express the Killing vector as
\begin{align}
  {\xi^\mu} = \alpha n^\mu + \beta^\mu \,.
\end{align}
For a stationary and axisymmetric spacetime, the Killing vector field can be
expressed as~\cite{Gourgoulhon2010}
\begin{align}
  {\xi^\mu} = \alpha n^\mu + \Omega \chi^\mu \,,
\end{align}
where $\chi^\mu$ is the rotational Killing vector field. In the case of a
perfect fluid that is not rotating or is experiencing circular flows, the
four-velocity of a zero angular momentum observer (ZAMO) is given by
\begin{align}
  {u^\mu} &= W \left[ n^\mu
          +  U^\mu \right]\,, \label{equ:fourvel}\\
  {U^\mu} &\equiv \frac{1}{\alpha} \left( \Omega \chi^\mu + \beta^\mu \right)\,,  \label{equ:3velgeneric} \\
  {W} &\equiv \frac{1}{\sqrt{1 - \gamma_{\mu\nu} U^\mu U^\nu}} \label{equ:Lorentz}\,,
\end{align}
where $W$ is the Lorentz factor with respect to a ZAMO, $U^\mu$ the fluid
3-velocity measured by a ZAMO, $\Omega$ is the angular velocity profile
of the fluid, and $\chi^\mu = \partial_\varphi$ is the rotation Killing
vector field for an axisymmetric spacetime.

\subsection{Quasi-isotropic coordinates}
A common choice of coordinates in numerical relativity are
quasi-isotropic coordinates (QIC) $\left(t\,, r\,, \theta\,, \varphi\right)$,
which preserve the axisymmetric nature of
the system while allowing for non-spherically symmetric
solutions~\cite{Gourgoulhon2010,Grandclement2014}.  For QIC,
$\gamma_{ij} = \left(A^2\,, A^2\,, B^2 r^2 \sin^2\theta\right)$ and
$\beta^i = \left(0\,, 0\,, - \omega \right)$,
yielding the following spacetime metric
\begin{align}
  g_{\mu\nu} \, \D x^\mu \, \D x^\nu   = & - \alpha^2 \D t^2 + A^2 \left(
    \D r^2 + r^2 \D\theta^2 \right) \nonumber \\
        & + B^2 r^2 \sin^2\theta \left(\D\varphi + \beta^\varphi \D t\right)^2.
                \label{equ:qicmet}
\end{align}
With this coordinate choice and the use of $\omega=- \beta^\varphi$,
\eqref{equ:Hconst} and \eqref{equ:Mconst} can be rewritten in a form that
is more suitable for numerical solutions~\cite{Gourgoulhon2010,
Paschalidis2016};
\begin{widetext}
  \begin{subequations}
    \label{equ:SYS_QIC}
    \begin{align}
      {\Delta_3 \nu} &=
        \frac{B^2r^2\sin^2\theta}{2 \alpha^2} \partial \omega \partial \omega
        - \partial\nu \partial\l(\nu + \ln B\r) + 4\pi A^2 \l(E+S\r) \,, \label{e:nu} \\
      {\tilde{\Delta}_3\l(\omega r \sin\theta\r)} &=
        -16 \pi \frac{\alpha A^2}{B^2} \frac{P_\varphi}{r\sin\theta}
        + r\sin\theta\partial \omega \partial\l(\nu - 3\ln B\r) \,, \label{e:Nphi} \\
      {\Delta_2\l(\ln A  + \nu\r)} &=
        \frac{3B^2 r^2\sin^2\theta}{4 \alpha^2} \partial \omega \partial \omega - \partial \nu \partial \nu
        + 8 \pi A^2 S^\varphi_{\ \, \varphi} \,,  \label{e:A} \\
      {\Delta_2 \l[\l(\alpha B -1\r)r \sin\theta\r]} &=
        8 \pi \alpha A^2 B r \sin\theta \l(S^r_{\ \, r} + S^\theta_{\ \, \theta} \r) \,, \label{e:B} \\
      {\Delta_3} &\equiv \frac{\partial^2}{\partial r^2} + \frac{2}{r}\frac{\partial}{\partial r} + \frac{1}{r^2} \frac{\partial^2}{\partial\theta^2} + \frac{1}{r^2 \tan \theta}\frac{\partial}{\partial \theta} \nonumber \\
      {\tilde{\Delta}_3} &\equiv \Delta_3 - \frac{1}{r^2\sin^2\theta} \,, \nonumber \\
      {\Delta_2} &\equiv \frac{\partial^2}{\partial r^2} + \frac{1}{r}\frac{\partial}{\partial r} + \frac{1}{r^2} \frac{\partial^2}{\partial\theta^2} \,, \nonumber\\
      {\partial f \partial g} &\equiv \frac{\partial f}{\partial r} \frac{\partial g}{\partial r} + \frac{1}{r^2}\frac{\partial f}{\partial\theta}\frac{\partial g}{\partial\theta} \,, \nonumber
    \end{align}
  \end{subequations}
\end{widetext}
where $\nu \equiv \ln \alpha$, $\Delta_3$ ($\Delta_2$) is the 3-dimensional (
2-dimensional) flat Laplacian and $\partial f \partial g$ denotes the flat metric scalar
product of the gradients of $f$ and $g$.

To model the fluid using QIC, we express the fluid 3-velocity \eqref{equ:3velgeneric} as~\cite{Gourgoulhon2010}
\begin{align}
  {U^\mu} &\equiv \frac{1}{\alpha} \left( \Omega  - \omega \right) \chi^\mu \,, \label{equ:3velQIC}
\end{align}
where $\omega$ is proportional to the angular velocity of a ZAMO due to
frame dragging effects of a rotating spacetime. Using the definition~\cite{Gourgoulhon2010}
\begin{align}
  {U} &\equiv \frac{B}{\alpha} \left( \Omega  - \omega \right) r \sin \theta \,, \label{equ:QIC_U}
\end{align}
and the fact that $\chi \cdot \chi = B^2 r^2 \sin^2\theta$, the fluid
3-velocity has the following $\varphi$ components~\cite{Gourgoulhon2010}
\begin{align}
  {U^\varphi} &= \frac{1}{\alpha} \left( \Omega  - \omega \right) \,, \\
  {U_\varphi} &= U B r \sin \theta \,. \\
\end{align}
With these definitions, we now express the Lorentz factor and
energy-momentum tensor projections $E$, $S$, $S^i_{\ \, j}$ and
$P_\varphi$ \eqref{equ:SYS_T} as~\cite{Gourgoulhon2010}
\begin{align}
  {W^2} &= \frac{1}{\left( 1 - U^2 \right)} \,, \label{equ:QIC_W} \\
  {E} &= \rho h W^2 - p \,, \label{equ:QIC_Energy} \\
  {P_\varphi} &= (E + p) U B r \sin \theta \,, \label{equ:QIC_Pphi} \\
  {S^r_{\hspace{2mm}r}} &= S^\theta_{\hspace{2mm}\theta} = p \,, \label{equ:QIC_Srrtt} \\
  {S^\varphi_{\hspace{2mm}\varphi}} &= p + \left( E + p \right) U^2 \,, \label{equ:QIC_Sphi} \\
  {S} &= 3p + \left( E + p \right) U^2 \label{equ:QIC_S} \,.
\end{align}

\subsection{eXtended Conformal Thin Sandwich (XCTS)}
In the case of compact objects with source terms, especially those that
include magnetic fields or in compact configurations such as binary
systems, the underlying degrees of freedom of the system are not readily
apparent. An incredibly successful ansatz is that of a conformal
decomposition of the spatial metric \cite{Lichnerowicz44,York73, York99}
such that
\begin{equation}
  {\gammaij} \equiv \Psi^4 \confgammaij \,, \label{equ:confgamma}
\end{equation}
where $\confgammaij$ is a background metric and $\Psi$
is the conformal factor that describes how the local region deviates from
the background metric. The extrinsic curvature can also be decomposed
into a trace free form~\cite{York99}
\begin{align}
    {K_{ij}} &= A_{ij} + \frac{1}{3} K \gammaij \,, \label{equ:Ktracefree} \\
    {A_{ij}} &= \Psi^{-10} \hat{A}_{ij} \,, \label{equ:confK} \\
    {\hat{A}^{ij}} &\equiv \frac{\Psi^6}{2 \alpha} (\tilde{D}^i \beta^j + \tilde{D}^j
        \beta^i - \frac{2}{3} \tilde{\gamma}^{ij} \tilde{D}_k \beta^k
            - \partial_t \tilde{\gamma}^{ij}) \,,
        \label{equ:SYS_Aij}
\end{align}
where the exponent of $\Psi$ is chosen to be consistent with the computed
result from the momentum constraint (cf. \cite{Gourgoulhon2007},
Sec.~6.5.2 for a discussion). By utilizing \eqref{equ:confgamma},
\eqref{equ:confK}, and \eqref{equ:SYS_Aij} we can rewrite
\eqref{equ:Hconst} \& \eqref{equ:Mconst} into what is known as the
Conformal Thin Sandwich decomposition (CTS):
\begin{align}
  \tilde{D}^k \tilde{D}_k \Psi +
      \frac{1}{8} \Psi^{-7} \hat{A}_{ij} \hat{A}^{ij} + 2 \pi \Psi^5 E &\nonumber \\
              - \frac{1}{8} \tilde{R} \Psi
              - \frac{1}{12} K^2 \Psi^5 &= 0 \,, \label{equ:SYS_conf} \\
  \tilde{D}_j \left(
    \frac{\Psi^6}{\alpha} (\tilde{\mathbb{L}} \beta)^{ij} \right)
        + D_j \left(\frac{\Psi^6}{\alpha} \partial_t \gamma^{ij} \right) &\nonumber \\
        - \frac{4}{3} \Psi^6 \tilde{D}^i K - 16 \pi \alpha \Psi^4  P^i &= 0 \,,
            \label{equ:SYS_shift} \\
  (\tilde{\mathbb{L}} \beta)^{ij} \equiv \tilde{D}^i \beta^j
            - \tilde{D}^j \beta^i
                - \frac{2}{3} \tilde{\gamma}^{ij} \tilde{D}_k \beta^k.& \nonumber
\end{align}
The remaining degrees of freedom yet to be specified are the conformal
metric $\confgammaij$ and its time derivative; the mean curvature $K$
and its time derivative; the lapse function $\alpha$; and the source
terms $E$ and $P^i$. In this work we restrict ourselves to the
conformally flat approximation such that conformal spatial metric in
Cartesian coordinates is $\confgammaij = \delta_{ij}$ and $\partial_t
\confgammaij = 0$. Furthermore, we assume maximal slicing for
adjacent hypersurfaces such that $K = \partial_t K = 0$.
An extension to the CTS system was made in~\cite{Pfeiffer:2005} where an
elliptic constraint equation is obtained for the lapse yielding
\begin{align}
  \tD_i \tD^i (\alpha \Psi)
      &+ \Psi^5(\partial_t K - \beta^i \tD_i K) \label{equ:SYS_lapse} \\
      &- \alpha \Psi \bigg[ \frac{1}{8} \tilde{R}
          + \frac{5}{12} K^2 \Psi^4
          + \frac{7}{8} \Psi^{-8} \hat{A}^{ij}\hat{A}_{ij} \nonumber \\
          & \hspace{1.6cm}+ 2 \pi \Psi^4 (E + 2S)\bigg]
               = 0. \nonumber
\end{align}
The complete system \eqref{equ:SYS_conf}, \eqref{equ:SYS_shift}, and
\eqref{equ:SYS_lapse} is commonly referred to as the eXtended Conformal
Thin Sandwich (XCTS) system of equations. Under our demand of conformal
flatness and maximal slicing, \eqref{equ:SYS_conf},
\eqref{equ:SYS_shift}, and \eqref{equ:SYS_lapse} take the form
\begin{subequations}
  \label{equ:SYS_xcts}
  \begin{eqnarray}
  \tilde{D}^k \tilde{D}_k \Psi +
  \frac{1}{8} \Psi^{-7} \hat{A}_{ij} \hat{A}^{ij}
      + 2 \pi \Psi^5 E &= 0 \,, \label{equ:SYS_Rconf} \\
  \tilde{D}_j \left(
      \frac{\Psi^6}{\alpha} (\tilde{\mathbb{L}} \beta)^{ij} \right)
          - 16 \pi \alpha \Psi^4 P^i &= 0 \,,
              \label{equ:SYS_Rshift} \\
  \tD_i \tD^i (\alpha \Psi) - \alpha \Psi \bigg[
  \frac{7}{8} \Psi^{-8} \hat{A}^{ij}\hat{A}_{ij} \nonumber \\
      + 2 \pi \Psi^4 (E + 2S)\bigg] &= 0 \,, \label{equ:SYS_Rlapse}
\end{eqnarray}
\end{subequations}
which are the actual equations implemented in the \fuka initial data solvers.
To obtain the source terms $E$, $S$, and $P^i$, we first find the fluid
3-velocity $U^i$ and Lorentz factor $W$ in Cartesian coordinates using
\eqref{equ:3velgeneric} and \eqref{equ:Lorentz}
\begin{align}
  {U^i} &= \frac{1}{\alpha} \left( \beta^i + \Omega \xi^i \right) \,, \\
  {W} &= \left( 1 - \Psi^4 U^i U_i \right)^{-1/2} \,.
\end{align}
With these definitions, we can compute the source terms \eqref{equ:SYS_T}
\begin{subequations}
  \label{equ:hydro_terms}
  \begin{align}
      {E} &= \rho h W^2 - p \,, \\
      {S^i_{\hspace{2mm}i}} &= 3p + (E+p) U^2 \,, \\
      {P^i} &= \rho h W^2 U^i \,.
  \end{align}
\end{subequations}

\subsection{Hydrostatic equilibrium}

To solve for a neutron star in hydrostatic equilibrium, we
enforce the conservation equations (see \cite{Papenfort2021b} for a
detailed discussion)
\begin{align}
  {\nabla_\mu T^{\mu \nu}} &= 0 \,, \label{equ:cons_Tmunu}\\
  {\nabla_\mu (\rho u^\mu)} &= 0. \label{equ:cons_denscurrent}
\end{align}
By expanding \eqref{equ:cons_Tmunu} for an isentropic fluid, we obtain
\begin{equation}
  u^\mu \nabla_\mu (h u_\nu) + \nabla_\nu h = 0. \label{equ:4euler}
\end{equation}
For a stationary spacetime where the fluid is undergoing stationary,
circular motion, the first integral of motion reduces
to~\cite{Gourgoulhon2010,Cassing2024}
\begin{align}
  \nabla_\mu \left[ \ln \left( \frac{\alpha h}{W} \right) \right] + j \nabla_\mu \Omega  &= 0 \,, \label{equ:firstint}\\
\end{align}
where $j \equiv u^t u_\varphi$ and is defined for QIC and XCTS coordinates as
\begin{align}
  {j_{\rm QIC}} & = \frac{W^2}{\alpha} U_\varphi \,, \\
  {j_{\rm CART}} & = \frac{W^2}{\alpha} \Psi^4 U_i \chi^i \,,
\end{align}
In the case of rigid rotation, \eqref{equ:firstint} reduces to
\begin{align}
  \ln \alpha +  \ln h - \ln W &= C \,, \label{equ:firstint_rigid}
\end{align}
where $C$ is a constant of integration.  However, in the case of
differential rotation, \eqref{equ:firstint} must be solved by determining
the integral of the fluid rotation profile, $j \nabla_\mu
\Omega$~\cite{Gourgoulhon2010, Cassing2024}. Within \fuka, we have
included the differential rotation model of \textit{Komatsu, Eriguchi,
and Hachisu} (KEH), also known as the $j$-constant model
\begin{align}
  {\Omega} &= \Omega_c - \frac{j}{\mathcal{A}^2} \,,
\end{align}
where $\mathcal{A}$ is a parameter that controls the degree of
differential rotation and $\Omega_c$ is the rotation velocity at the
stellar center. In the limit of $\mathcal{A} \to \infty$, the KEH model
reduces to a rigidly rotating star. In order to find a solution for
$\mathcal{A}$ and $\Omega_c$, we must also specify the ratio of the polar
to equatorial radius of the star ($r_p / r_e$) along with the ratio
$\hat{\mathcal{A}} \equiv \mathcal{A} / r_e$. For this model, an
analytical first integral is available for the rotation profile resulting
in a first integral of motion of the form

\begin{align}
  \ln \alpha +  \ln h - \ln W - \frac{1}{2} \frac{j^2}{\mathcal{A}^2}  &= C \,. \label{equ:firstint_KEH}
\end{align}

To close the system of equations, an equation of state (EOS) must also be
specified. Within \fuka one can utilize (piecewise--)polytropic
EOSs, cold tabulated EOSs in the standard \texttt{LORENE} format, or 3D
tabulated EOSs in \textit{stellar collapse} format~\cite{stellarcollapse}. Support for 3D
tabulated EOS is provided by the \grhayl library~\cite{Cupp2025, GRHayL},
which has been recently incorporated into \fuka. We note that
the current use of \grhayleos is restricted to the cold
beta-equilibrium slice of the 3D EOS table. The reason for this is
twofold: (i) the EOS module in \fuka assumes an isentropic fluid ($ds=0$)
and (ii) \grhayleos only supports computing constant temperature,
beta-equilibrium slices of the EOS table. Future work will include
support for constant entropy slices of the EOS table as well as removing
the assumption of an isentropic fluid in \fuka to enable the
self-consistent study of hot stars.

\subsection{Diagnostics}
To reliably fix the characteristics of the NS, we need to compute
specific diagnostic quantities. To fix the mass of the NS, we
either fix fluid quantities at the center of the star (e.g.,
central rest-mass density or central specific enthalpy) or the
total gravitational mass of the NS. The latter is fixed by computing
the ADM mass of the system. In \fuka, this is performed by computing a
surface integral on the outer boundary of the outermost domain.  This
domain leverages a compactified coordinate system such that $r =\infty$
is mapped to a finite value~\cite{Grandclement2009}.  In this way, we can
reliably compute the ADM mass in the weak field limit using
\begin{align}
  {M^{^{\rm QIC}}_{\rm ADM}} &= - \frac{1}{4 \pi} \int_{S_\infty} \partial_i B \D s^i  \,, \\
  {M^{^{\rm XCTS}}_{\rm ADM}} &= - \frac{1}{2 \pi} \int_{S_\infty} D_i \Psi \D s^i  \,.
\end{align}
As a consequence of fixing the mass of the star using the ADM mass or
central fluid quantities, it is possible to compute the total rest-mass
of the star by computing the following volume integral over the stellar
domains:
\begin{align}
  {M_{\rm b}} &= \int_{V} W \rho \sqrt{\gamma} \D x^3   \,.
\end{align}

Additional diagnostics beneficial for analyzing the ID solutions and
comparing to previous literature include the total internal energy
$E_{\rm int}$, the total angular momentum $J$, the rotational kinetic
energy $T$, the gravitational binding energy $\mathcal{W}$ and the
instability parameter, $\beta$\footnote{We use $\beta$ to denote the
instability parameter to be consistent with previous literature, however;
it should not be confused with the shift vector,
$\beta^i$.}~\cite{Shibata2000, Loffler2014}
\begin{align}
  {E_{\rm int}} &= ~~\int_{V} W \rho \epsilon \sqrt{\gamma} \D x^3 \,, \\
  {J} &= ~~\int_{V} \rho j \alpha \sqrt{\gamma} \D x^3 \,, \\
  {T} &= \frac{1}{2} \int_{V} \Omega \rho j \alpha \sqrt{\gamma} \D x^3 \,, \\
  {\mathcal{W}} &= T + E_{\rm int} + M_b - M_{\rm ADM} \,, \\
  {\beta} &= \frac{T}{\left| \mathcal{W} \right|} \,.
\end{align}

\subsection{Boundary Conditions}
To close the system of equations, appropriate boundary conditions must be
imposed. At the outer boundary of the computational domain, we impose the
weak-field conditions on metric variables, such that for XCTS:
\begin{align}
  {\lim_{r \to \infty} \alpha} &= 1 \,, \\
  {\lim_{r \to \infty} \Psi} &= 1 \,, \\
  {\lim_{r \to \infty} \beta^i} &= 0 \,,
\end{align}
and for QIC:
\begin{align}
  {\lim_{r \to \infty} \nu} &= 0 \,, \\
  {\lim_{r \to \infty} (\ln A + \nu)} &= 0 \,, \\
  {\lim_{r \to \infty} (\alpha B - 1)r \sin\theta} &= 0 \,, \\
  {\lim_{r \to \infty} (\omega r \sin\theta)} &= 0 \,.
\end{align}

To ensure the surface of the star corresponds to a domain boundary, we
set the boundary condition for the adapted domain shells to be defined by
\begin{align}
  \ln h &= 0 \,.
\end{align}

\section{Implementation Details}
\label{sec:impl}
The \fuka suite of initial data codes is built on an extended version of
the \kadath spectral library~\cite{Grandclement2009}, a robust framework
designed for solving elliptic systems of equations using spectral
methods. The relevant features of \kadath critical to this work include its novel
interface for implementing equations and constraints in a
\LaTeX-like syntax, which are then automatically converted into numerical
algorithms that are solved using spectral methods.

\kadath utilizes a novel domain decomposition that adopts
a nucleus domain with a well-defined coordinate system across and including $r
= 0$, adapted domain shells that allow for non-spherical/circular objects
(i.e., rotating neutron stars) and exterior domain shells that use
compactified coordinates such that $r = \infty$ (i.e., the weak-field limit)
maps to a finite value.
Use of adapted shells is essential for numerical stability,
so that the stellar surface can be defined at the grid interface.
This results in reduced spectral noise, known as
Gibbs phenomenon, which occurs when attempting to model discontinuities
using a single spectral expansion.

Systems of equations within \kadath are solved using a Newton-Raphson
method, employing automatic differentiation to efficiently compute
derivatives of fields in parallel. The Newton-Raphson method is
used in conjunction with a tau method, to find spectral coefficients
that minimize the residual of the constraint
equations. In \fuka, a stopping criterion of $\varepsilon_{\rm STOP}
\equiv 10^{-8}$ is used.

\subsection{QIC Solver}

To compute solutions of the QIC system of equations\footnote{In the \fuka
repo, the solver is called \textit{NS\_isotropic}}, \fuka utilizes
\kadath's support for polar coordinate systems, which are incredibly
efficient for solving problems with circular axial symmetry (see
Ref.~\cite{Grandclement2014} for details). This reduces the dimensionality
of the problem to 2D, and significantly reduces the degrees of freedom.

The physical space is decomposed into numerical domains including a polar
nucleus domain that covers the center of the star followed by two adapted
polar shells that adapt to the stellar surface. An outer compactified
polar shell is used to impose the boundary conditions in the weak-field
limit. All domains are matched at domain interfaces where the boundary
conditions simply enforce continuity of the fields and their normal
derivatives.

To initialize the fluid and spacetime fields, \fuka utilizes a 1D
Tolman–Oppenheimer–Volkoff (TOV) solution of an isolated,
non-rotating star using the same EOS and target
mass. If the desired mass is not supported by the TOV solver, the
maximum mass for the EOS ($M_{\rm TOV}$) is used and the desired mass is
attempted to be found once the desired rotation profile is found.
\fuka then solves the QIC system of equations \eqref{equ:SYS_QIC} in
conjunction with thermodynamic constraint equations \eqref{equ:firstint_rigid} for a
non-rotating star. Once a solution is found, a uniform rotation profile (e.g., $\chi = 0.1$)
is solved for. This is important, as the differential
rotation parameters only specify the shape of the star and the degree
of differential rotation, but not the direction. Therefore, the
Newton-Raphson solver will quickly diverge if attempting to solve for a
differentially rotating star directly.
Finally, once a uniformly rotating solution is found, the axis ratio
($r_p / r_e$) is incrementally decreased to the desired value for a fixed
value of $\hat{\mathcal{A}}$.

The simplicity of the QIC system combined with the efficiency of
\kadath's spectral solver enables the QIC solver to run on moderate
system resources (e.g., quad-core CPU and $\sim 16$ GB of computer
memory). Even at high resolutions, solutions can be computed in minutes,
and sequences of equilibrium solutions, such as those presented in Fig.~\ref{fig:rns_cmp},
take less than an hour to compute.

\subsection{XCTS Solver}

In the case of solving the 3D XCTS system of equations, \fuka utilizes
Cartesian coordinates for physical quantities within \kadath's spherical
domain decomposition, whereas a spectral expansion over $(r, \theta,
\varphi)$ for the radial, polar, and azimuthal directions are used for the
numerical coordinates.

The physical space is decomposed in a similar fashion to the QIC space, only now
3D spherical domains are used instead, i.e., a spherical nucleus domain, two
adapted spherical shells, and an outer compactified spherical shell. The
boundary conditions are imposed in a similar manner as the QIC solver for
the relevant fields.

While the XCTS system is a flexible system of equations and open to extension to a
variety of physical systems (e.g., binary initial data, misaligned spins,
magnetic fields, etc.,)~\cite{Papenfort2021b, Rashti2024, Rashti2021}, it
is computationally expensive. Beyond coarse resolution models, high
resolution solutions even for isolated neutron stars can take hours/days
to compute on a personal computer with sufficient memory.

As we will demonstrate in Sec.~\ref{sec:results}, the QIC and XCTS
solutions are in excellent agreement and approximately recover the same
physical metric and fluid quantities. Therefore, we utilize a hybrid
approach to compute NS solutions using the XCTS system of equations in
the publicly available codes\footnote{In the \fuka repo, the solver is
called \textit{NS\_DIFFROT}}. To do so, we first compute the desired
solution using the QIC solver. Using \fuka's export interface for QIC
solutions, we obtain the physical metric and fluid quantities in
Cartesian coordinates. Next, we initialize the XCTS fields by
interpolating the QIC solution at the Cartesian coordinates of the XCTS
collocation points. Here we approximate the conformal factor as $\Psi^4 =
\gamma_{xx}$. Once the fields are initialized, \fuka will solve the XCTS
system of equations \eqrefs{equ:SYS_Rconf}{equ:SYS_Rlapse} and
thermodynamic constraints for the desired solution directly
(non-rotating, uniformly rotating, or differentially rotating) based on
the QIC solution, thus significantly reducing the computational time to
find an XCTS solution.

\subsection{Resolution}
The spectral resolution in \fuka is currently specified by the user using
a single quantity, \textit{res} ($R$). For QIC solutions, \textit{res}
specifies the number of collocation points in $r$ and $\theta$ directions
per domain, $(R,R)$. For XCTS solutions, \textit{res} specifies the
number of collocation points in $r$, $\theta$, and $\varphi$ directions per
domain, $(R,R,R-1)$. While angular resolution plays an important role in
resolving the deformed star, the radial resolution is essential for
comparing ID solutions and understanding when spectral noise begins to deteriorate
the solution.

An additional quantity that will prove useful in Sec.~\ref{sec:results} is
the effective spectral resolution, $\tilde{N}$, which we define as
\begin{align}
  {\tilde{N}} &\equiv (D/3)N^{1/D} \,, \label{eq:effective_resolution} \\
  {N} &\equiv \sum_{i=1}^{d} \prod_{j \in {r \,, \theta \,, \varphi}} N^{(i)}_j \,,
\end{align}
where $N^{(i)}_j$ is the number of collocation points in the $j$-th
direction of the $i$-th domain, $d$ is the number of domains, and $D$ is
the number of spatial dimensions (i.e., $D=2$ for QIC and $D=3$ for XCTS).
The effective spectral resolution provides an estimate of the average
number of collocation points per direction across all domains and
provides an additional means to effectively compare QIC and XCTS
solutions in Sec.~\ref{sec:results}.

\section{Methods --- Dynamic Simulations}
\label{sec:methods}

\begin{table*}[ht]
  \caption{Comparison of the \sbsix and \uteen models between XCTS and
    QIC solvers at resolutions $R \in \{11, 13, 17\}$ against reference
    values from Ref.~\cite{Baiotti2006}. Here we list the instability
    parameter $\beta$, central density $\rho_c$, and dimensionless
    angular momentum $J/M^2$ for each model. \sbsix is fixed using $r_p /
    r_e = 0.336$, $\hat{\mathcal{A}} = 1$, and $M_{\rm ADM} = 1.449$.
    \uteen is fixed using $r_p / r_e = 0.20012$, $\hat{\mathcal{A}} = 1$,
    and $M_{\rm ADM} = 1.462$. \sbsix and \uteen utilize a polytropic EOS
    with $\Gamma = 2$ and $K = 100$ whereas \ddsbsix and \dduteen
    utilize the \texttt{SLy} 3D tabulated EOS.}
  \setlength{\tabcolsep}{16pt}
  \label{tab:gam2_models}
  \begin{longtable*}[t]{l|ccc|lccc}
    \hline
    \textbf{\sbsix} & \textbf{$\beta$} & \textbf{$\rho_c \left[10^{4}\right]$} & \textbf{$J/M^2$} &
    \textbf{\uteen} & \textbf{$\beta$} & \textbf{$\rho_c \left[10^{5}\right]$} & \textbf{$J/M^2$}
    \\ \hline \hline
    \endfirsthead
    Ref.~\cite{Baiotti2006} & 0.240 & 2.261 & 1.411 &
    Ref.~\cite{Baiotti2006} & 0.281 & 5.990 & 1.753
    \\
    XCTS.R11 & 0.236 & 2.261 & 1.410 &
    XCTS.R11 & 0.285 & 5.983 & 1.738 \\
    XCTS.R13 & 0.236 & 2.261 & 1.411 &
    XCTS.R13 & 0.278 & 5.995 & 1.791 \\
    XCTS.R17 & 0.236 & 2.261 & 1.410 &
    XCTS.R17 & 0.279 & 5.992 & 1.760 \\
    QIC.R11  & 0.234 & 2.263 & 1.410 &
    QIC.R11  & 0.276 & 5.959 & 1.746 \\
    QIC.R13  & 0.234 & 2.263 & 1.413 &
    QIC.R13  & 0.274 & 6.008 & 1.754 \\
    QIC.R17  & 0.234 & 2.263 & 1.413 &
    QIC.R17  & 0.274 & 5.996 & 1.754 \\ \hline
    \textbf{\ddsbsix} & & &  &
    \textbf{\dduteen} & & \textbf{$\rho_c \left[10^{4}\right]$} &
    \\ \hline \hline
    XCTS.R11 & 0.226 & 6.819 & 1.136 &
    XCTS.R11 & 0.276 & 3.510 & 1.449 \\
    XCTS.R13 & 0.228 & 6.797 & 1.133 &
    XCTS.R13 & 0.266 & 3.484 & 1.315 \\
    XCTS.R17 & 0.229 & 6.801 & 1.134 &
    XCTS.R17 & 0.272 & 3.490 & 1.349 \\
    \hline
  \end{longtable*}
\end{table*}

To validate initial data solutions computed using \fuka, we must examine
the impact of the ID solution on the dynamical behavior of the NSs after
evolution. To this end we utilize the public evolution framework, the
\etk~\cite{EinsteinToolkit:2025_05}, and the fixed-mesh box-in-box
refinement driver~\cite{Schnetter2003}. Within the \etk, we make use of
the general relativistic magnetohydrodynamics (GRMHD) code
\igm~\cite{Etienne2015} to evolve the hydrodynamic quantities coupled to
the spacetime evolution provided by the \texttt{Baikal} thorn.

Baikal employs the BSSN formulation~\cite{Baumgarte1998, Nakamura1987,
Shibata1995}, where we specify the use of fourth order finite
differencing, and employ the fourth-order Runge-Kutta timestepping
algorithm to evolve the coupled matter-spacetime system, as implemented
in the \etk, using a Courant factor of $0.2$. For the spacetime evolution
we also employ fifth order Kreiss-Oliger
dissipation~\cite{Kreiss1973MethodsFT}, with strength $0.2$, and use the
moving puncture gauge condition, with gamma-driving
shift~\cite{Campanelli2005,Baker:2005vv,Alcubierre:2002kk}. For outer
boundary conditions we use outgoing radiation boundary conditions using
the NewRad thorn, while \igm enforces outflow boundary conditions on the
fluid velocity.

Here we validate the ID solutions computed in \fuka by computing and
evolving polytropic models \sbsix and \uteen, which have been thoroughly
studied in the literature~\cite{Baiotti2006} (see
Tab.\ref{tab:gam2_models}). For these models we use \igm presented in
Ref.~\cite{Etienne2015} with the same equation of state parameters as the
initial data models. We note that \igm uses a hybrid equation of state of
the form
\begin{align}
  P = K\rho^\Gamma + \left(\Gamma_{\mathrm{th}} - 1 \right)\rho \epsilon_\mathrm{th},
\end{align}
where $\epsilon_\mathrm{th}$ is the thermal contribution to the internal
energy originating from shocks. Therefore we set $\Gamma_{\mathrm{th}} =
\Gamma$ in the tests presented here.

For each simulation we use a constant density atmosphere
$\rho_\mathrm{atm} = 1.28 \times 10^{-12}$ and exclude magnetic field
evolution, similar to the studies presented in \cite{Manca2007,
Loffler2014}.

Each configuration is evolved for $\sim 30$ms using two resolutions in
order to confirm convergence of the ID models in an evolution scheme. For
our medium resolution ($\Delta_M x$), we use only two refinement levels
centered on the neutron star, with the coarse grid having a resolution
$dx = 0.8$ with side-lengths $204.8$ in all three directions. In our
evolution of these models we do not impose any symmetries, and only allow
truncation errors to drive the bar-mode instability. For our high
resolution simulations ($\Delta_H x$) we decrease the grid spacing on
both refinement levels by 20\%, where the coarse grid has resolution $dx
= 0.64$.

We have chosen to use \sbsix (\uteen) models as they are stable
(unstable) against the dynamical bar mode instability, where the NS
becomes deformed into a bar-like configuration. During this transient
period, the NS expels matter and redistributes angular momentum, until
relaxing into a new stable configuration. The study presented in
\cite{Loffler2014} demonstrated that models of rapidly rotating NS with a
rotation profile described by the KEH law will be susceptible to the
instability if the $\beta = T/\left| W \right|$ parameter exceeds some
critical value. To track the growth and saturation of the instability
during dynamical evolution, we monitor the azimuthal Fourier modes using
\begin{align}
  P_m = \int_V d^3x \rho e^{im\varphi}.
\end{align}
The bar-mode instability is characterized by an exponential increase in
the $m=2$ mode~\cite{Baiotti2006}. To this end, we measure $P_m$ out to a radius of
$r=45$ throughout the time evolution.

Finally, we aim to validate solutions computed with \fuka using a
finite-temperature, tabulated equation of state, using the same methods
for dynamical evolution only now with the SLy EOS~\cite{Chabanat1997}.
Since existing literature has not focused on the bar-mode instability
with realistic EOS, we construct similar models to the polytropic cases
presented above, such that a stable and unstable model are obtained (see
Tab.~ \ref{tab:gam2_models}), however, we focus solely on the unstable
model computed with the XCTS solver is this proved sufficient to validate
the solutions computed with \fuka. For these tests we utilize the
tabulated EOS version of \igm as described in~\cite{Werneck2022}, with
the electron fraction within the star initialized based on the
beta-equilibrium cold slice.

\section{Results}
\label{sec:results}

Here we present the initial data and time evolution results using QIC and
XCTS solutions. We will begin with a self-consistency test of the XCTS
and QIC solvers independently followed by a comparison of the agreement
between the two solvers. With confidence in the equivalence of the two
solvers, we will compare the QIC solution against the \hrns thorn from
the \etk, derived from the \rns code~\cite{Stergioulas95, Stergioulas04}
for a variety of differential rotation profiles. Additionally, we will
present a rigorous analysis of the time evolution of the QIC and XCTS
solutions for configurations that are bar-mode (un)stable and compare to
previously published results.  Finally, we will present novel results
using dynamical simulations of a bar-mode unstable configuration using a
realistic, tabulated EOS.

For the results presented using a polytropic EOS (see
Tab.~\ref{tab:gam2_models}), we utilize the \sbsix and \uteen
configurations which have been thoroughly studied
previously~\cite{Baiotti2006}. In this case, the EOS is a simple
polytrope with a polytropic index of $\Gamma = 2$ and a polytropic
constant of $K = 100$.

\begin{figure}[t]
  \includegraphics[width=\columnwidth]{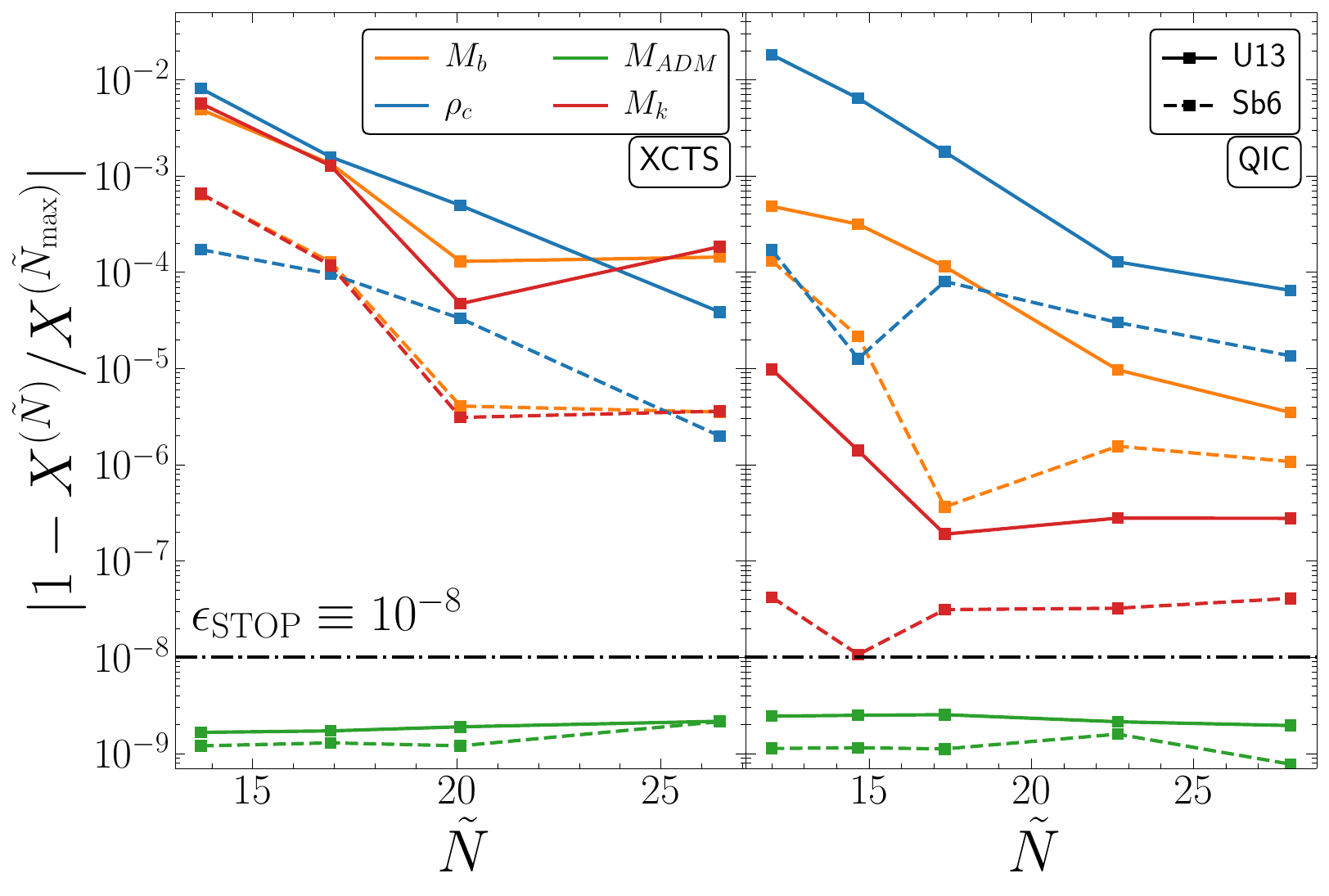}
  \caption{ Self-convergence of the XCTS (\textit{left}) and QIC
    (\textit{right}) solvers for \uteen and \sbsix
    configurations, respectively. Exponential convergence is observed up
    to $\tilde{N} \sim 22$ at which point the error is dominated by
    spectral noise and solutions begin to deteriorate. We include
    $\varepsilon_{\rm STOP}$, the stopping threshold for the
    Newton-Raphson solver in \kadath, to highlight the expected precision
    floor. }
  \label{fig:idconvergence}
\end{figure}

\subsection{Initial Data Convergence}
A key feature of using spectral methods is the ability to achieve
exponential convergence of the solution as the number of collocation
points is increased.
To verify this, we compute a series of solutions for both the \sbsix and
\uteen configurations at increasing resolutions using the XCTS and QIC
solvers. For a given effective resolution $\tilde{N}$ as defined using
\eqref{eq:effective_resolution}, we compute the ADM mass ($M_{ADM}$), the
Komar mass ($M_K$), the baryonic mass ($M_b$), and the central rest-mass
density ($\rho_c$).  Finally, we compute the relative difference of each
quantity with respect to the highest resolution solution for each
configuration and solver.  In Fig.~\ref{fig:idconvergence}, we illustrate
the self-convergence of the \sbsix (dashed line) and \uteen (solid)
configurations, using the XCTS (left) and QIC (right) solvers. Included
in the plot is a horizontal line denoting the stopping threshold of the
Newton-Raphson solver in \fuka, $\varepsilon_{\rm STOP} \equiv 10^{-8}$.

Both XCTS and QIC solutions exhibit the expected exponential convergence
up to $\tilde{N} \sim 22$ at which point the error is dominated by
spectral noise and the solution begins to deteriorate.  We find that the
QIC solver exhibits considerably better convergence properties compared
to the XCTS solver. This is primarily attributed to the reduced spectral
noise given the spectral bases for QIC is only a product of two
coordinate bases rather than three for XCTS. Additionally, the reduced
dimensionality and computational complexity of the QIC system also
contributes to the improved convergence properties. Finally, we note that
the ADM mass, which is used as a fixing parameter for each configuration,
lies well below $\varepsilon_{\rm STOP}$, indicating that the solvers are
able to find a solution that is consistent with the desired mass of the
configuration.

\begin{figure}[t]
  \includegraphics[width=\columnwidth]{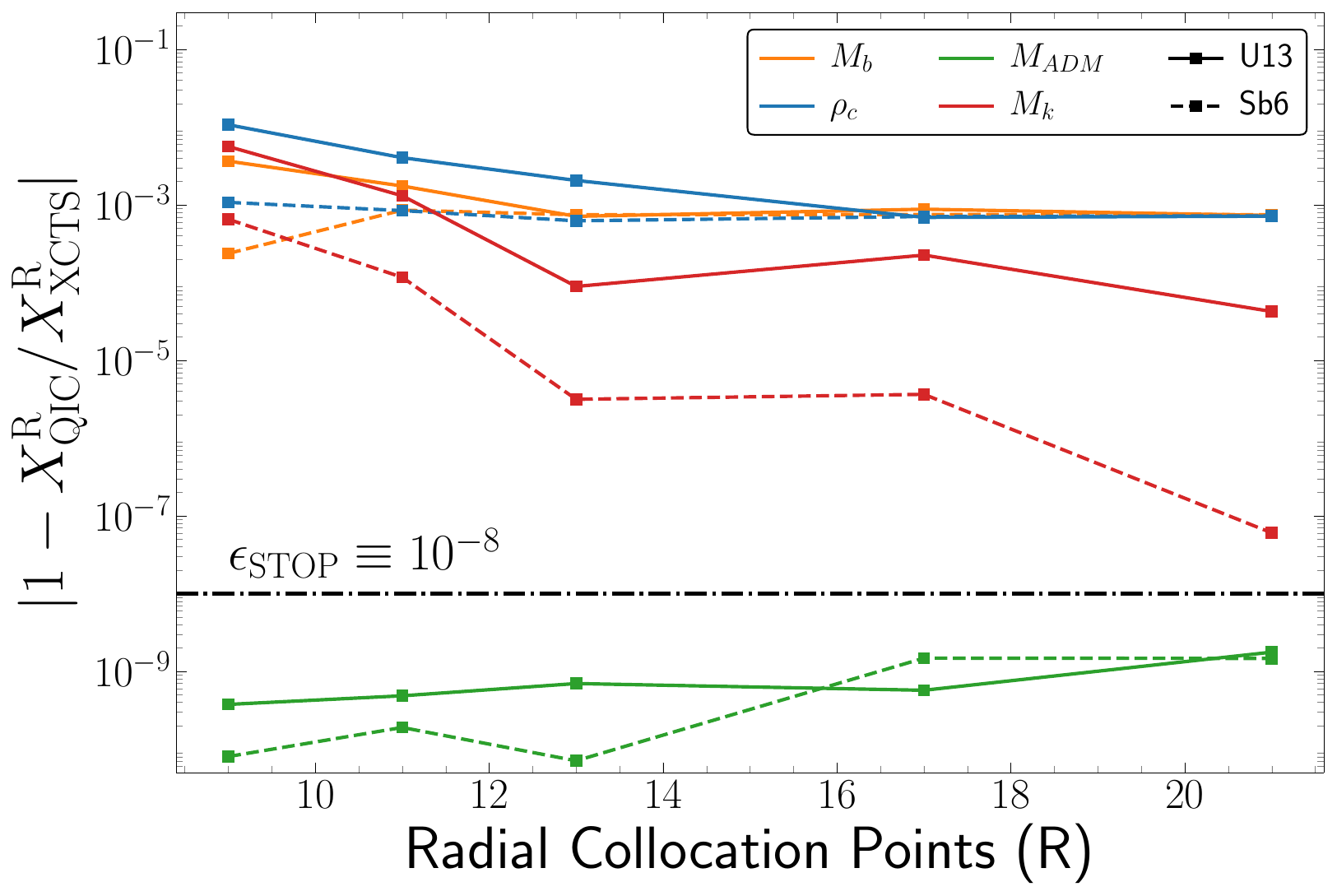}
  \caption{ Comparison between diagnostic quantities computed using QIC
    and XCTS solutions for a range of radial collocation points $R$.  We
    find excellent agreement of $M_K$ and $M_{ADM}$ as the radial
    resolution is increased, however, fluid quantities such as $M_b$ and
    $\rho_c$ only show convergence up to $10^{-3}$. }
  \label{fig:solution_cmp}
\end{figure}

To compare the solutions obtained from the QIC solver against those from
the XCTS solver, we use instead the radial collocation point resolution
$R$ instead of the effective resolution $\tilde{N}$ to make a more direct
comparison. Using the solutions computed previously, we calculate the
relative difference of the QIC and XCTS solutions for the same diagnostic
quantities as a function of $R$, which is illustrated in
Fig.~\ref{fig:solution_cmp}. While we find excellent agreement of the ADM
and Komar masses, the fluid quantities, $M_b$ and $\rho_c$, only agree to
$\sim 10^{-3}$.  This points to either an inaccurate comparison due to
differences in effective resolution or systematic differences due to the
underlying numerical methods and system of equations. Overall, we conclude
that the solutions computed by the XCTS and QIC
solvers are consistent and achieve the desired accuracy.

Finally, we list in Tab.~\ref{tab:gam2_models} the key diagnostic
quantities for the \sbsix and \uteen configurations computed using both
the XCTS and QIC solvers for the resolutions considered, alongside the
reference values from Ref.~\cite{Baiotti2006}. Overall we find excellent
agreement between both solvers and the reference values on the order of
three significant figures.

\begin{figure}[t]
  \includegraphics[width=\columnwidth]{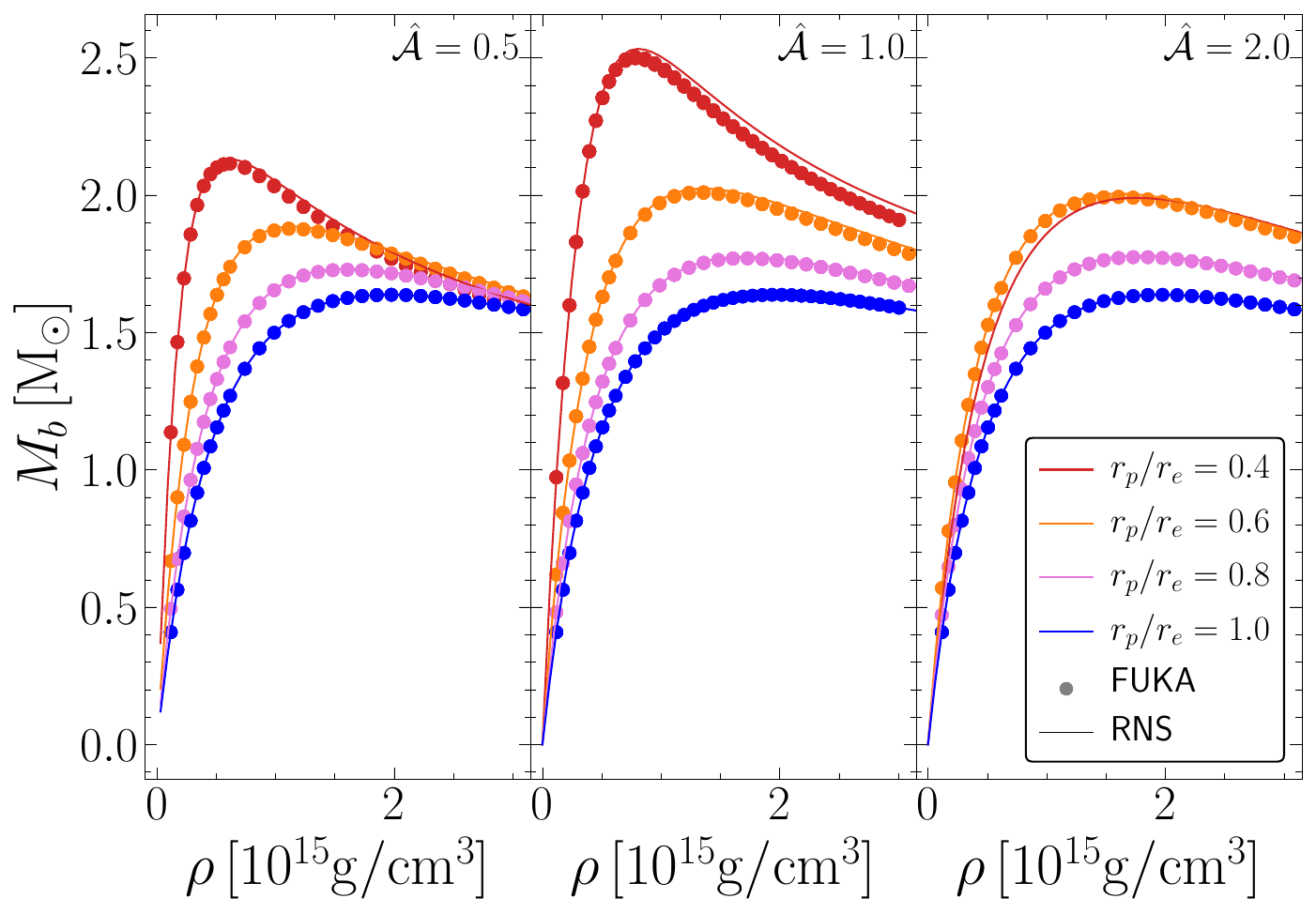}
  \caption{ Comparison of solutions between the \fuka QIC solver
  (circles) and the \rns code (solid lines). Here we explore a range of
  differential rotation profiles $\hat{\mathcal{A}} = \left\lbrace 0.5
  \,, 1.0 \,, 2.0 \right\rbrace$ (left, middle, right panes,
  respectively) and axis ratios $r_p / r_e = \left\lbrace 0.4 \,, 0.6 \,,
  0.8 \,, 1.0 \right\rbrace$ (red, orange, pink, and blue lines,
  respectively) over a range of central densities. No solutions were
  found with \fuka for $\hat{\mathcal{A}} = 2.0$ and $r_p / r_e = 0.4$,
  but the \rns solution is included for completeness and to highlight the
  change in behavior in this region of the parameter space.}
  \label{fig:rns_cmp}
\end{figure}

\subsection{Comparison with RNS}
With the confidence that the QIC solutions are self-consistent and agree
with the XCTS solution, we now compare the QIC solutions against those
computed using the well-established \rns code~\cite{Stergioulas95,
Stergioulas04}. In Fig.~\ref{fig:rns_cmp}, we plot the baryonic
mass ($M_b$) as a function of the central rest-mass density ($\rho_c$)
for a series of solutions for $\hat{\mathcal{A}} = \left\lbrace 0.5 \,,
1.0 \,, 2.0 \right\rbrace$ (left, middle, right panes, respectively) and
$r_p / r_e = \left\lbrace 1.0 \,, 0.8 \,, 0.6 \,, 0.4 \right\rbrace$
(blue, pink, orange, and red lines, respectively) over a range of central
densities. Here we denote the solution computed with \fuka using circle
markers, while solid lines indicate solutions computed with \rns.
For all configurations, a polytropic EOS is used with $\Gamma = 2$ and $K
= 100$.

\begin{figure}[t]
  \includegraphics[width=\columnwidth]{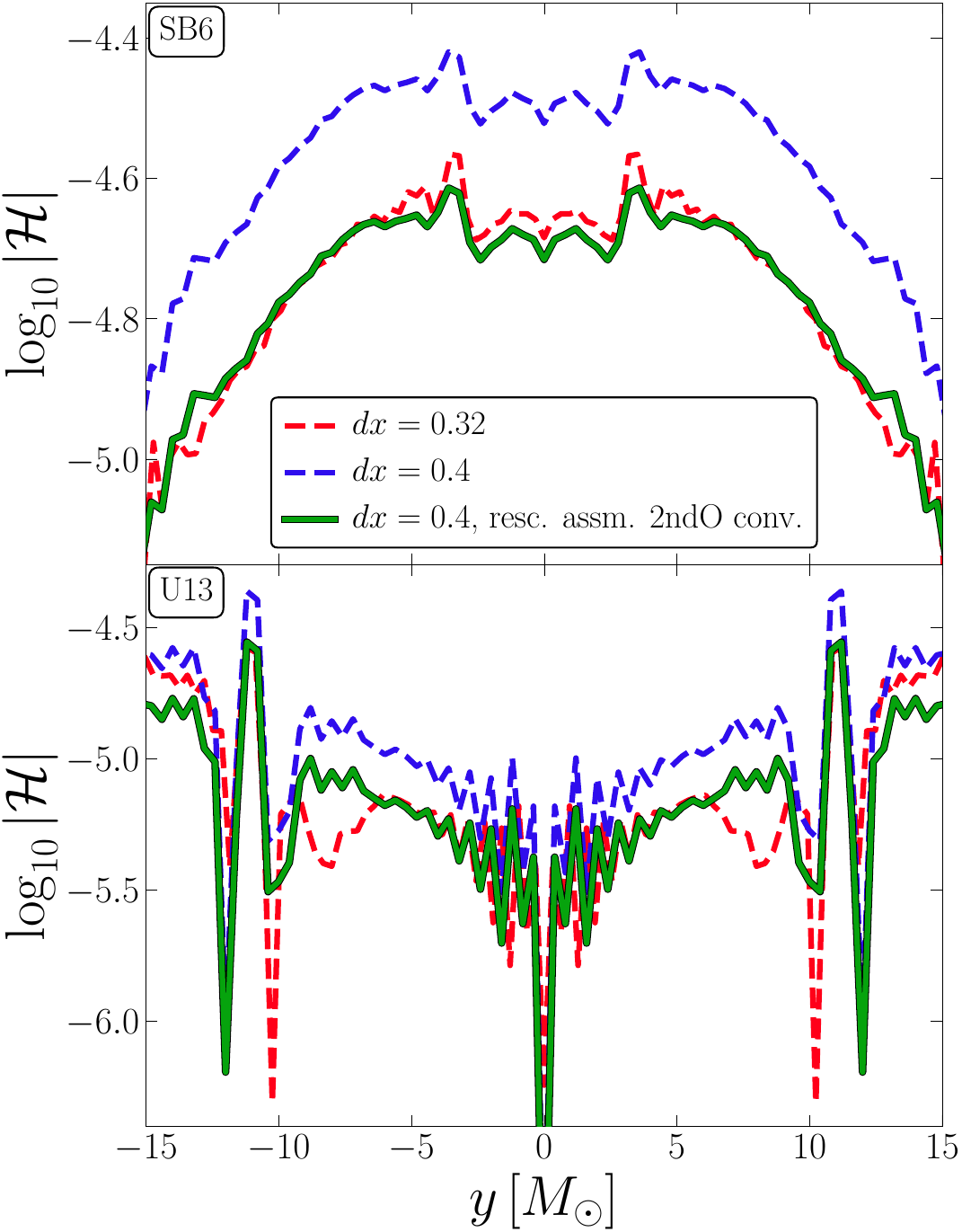}
  \caption{ Convergence analysis of the Hamiltonian constraint violations
  along the $x$-axis at $t = 41 M$ where we find the expected
  second-order convergence as predicted by the hydro evolution code,
  \igm.}
  \label{fig:evo_H_convergence}
\end{figure}

For $\hat{\mathcal{A}} = 0.5$ (left pane), we find excellent agreement
between the two codes for all values of $r_p / r_e$.  In the case of $r_p
/ r_e = 0.4$, we find that \fuka solutions become slightly less massive
than those computed with \rns. For $\hat{\mathcal{A}} = 1.0$, we find
excellent agreement for $r_p / r_e = 1.0$ and $0.8$ with slight
deviations for $r_p / r_e = 0.6$ and stronger deviations for $r_p / r_e =
0.4$. Finally, for $\hat{\mathcal{A}} = 2.0$ we find similar agreement for
$r_p / r_e = 1.0 \,, 0.8 \,, {\rm and} \,, 0.6$, but solutions for
$r_p / r_e = 0.4$ were not able to be computed using \fuka.  We note that
\rns solutions for $\hat{\mathcal{A}} = 2.0$ $r_p / r_e = 0.4$ show
interesting behavior where the stable masses are less than those for $r_p
/ r_e = 0.6$, which is in contrast to $\hat{\mathcal{A}} = 1.0 \,, 0.5$.
We leave this as an open question for potential future study in order to
ascertain the stability of solutions in this region of the parameter space
and its utility to model astrophysically relevant systems.

With the exception of $\hat{\mathcal{A}} = 2.0$ $r_p / r_e = 0.4$, we
find the solutions computed with \fuka are consistent and in agreement
with those computed with \rns. The slight deviations observed for the
most extreme systems exist on the unstable branch of the solution space
and, thus, are likely sensitive to the numerical methods and resolution
used to compute the solutions.

\begin{figure}[t]
  \includegraphics[width=\columnwidth]{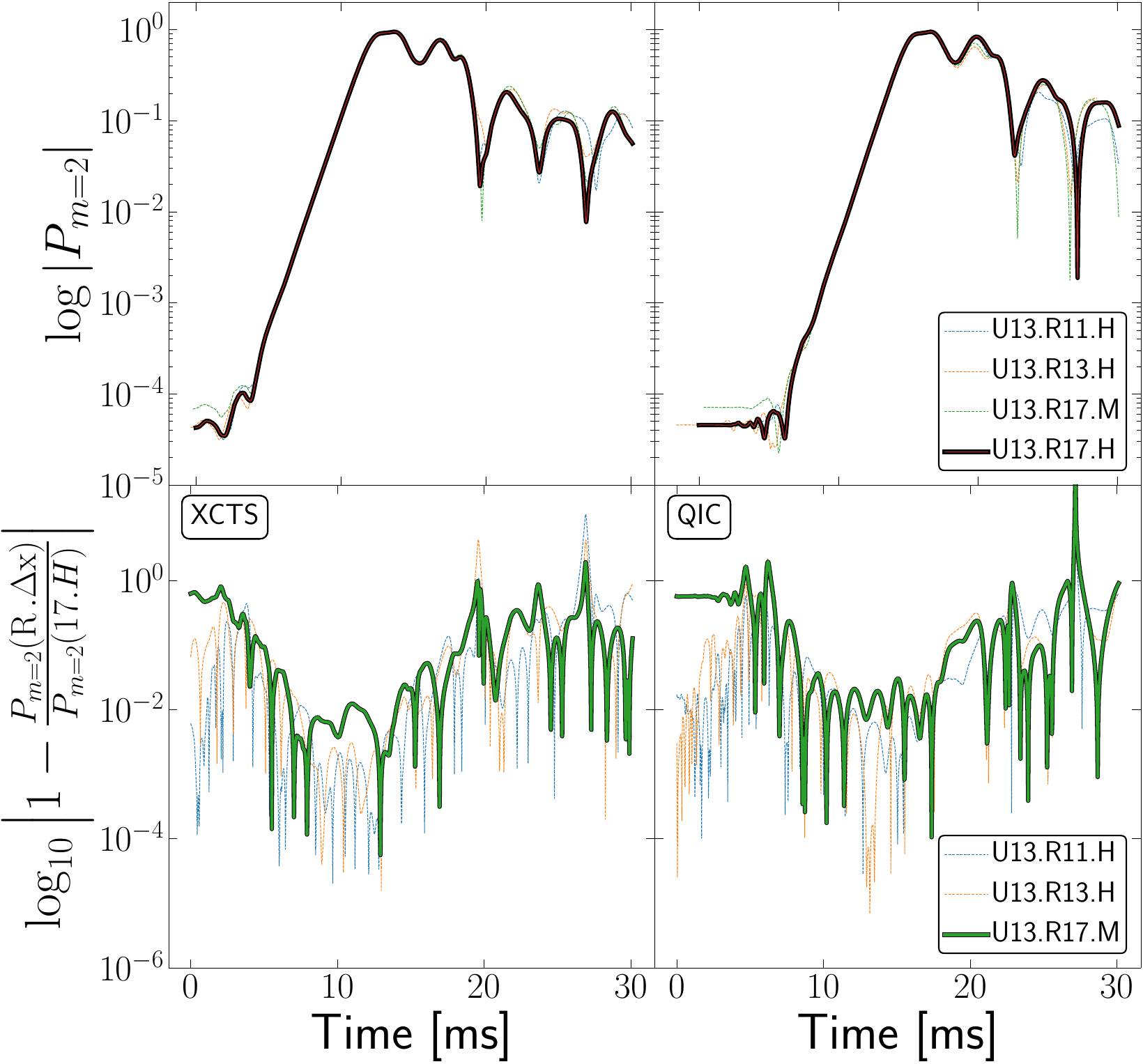}
  \caption{ \textit{Top}: Evolution of the $m = 2$ Fourier mode of the
  stellar interior for the XCTS solution (left) and QIC solution (right)
  for the \uteen configuration. Both are consistent with behavior
  reported in Ref.~\cite{Baiotti2006}. \textit{Bottom}: Convergence
  analysis of the $m = 2$ Fourier mode of the stellar interior during
  dynamical evolution for the \uteen configuration to assess the impact
  of initial data resolution $R = \left\lbrace11 \,, 13 \,,
  17\right\rbrace$ at evolution resolution $\Delta_H x = 0.32$ and
  $\Delta_M x = 0.4$. For the QIC results, we have aligned the datasets
  at the onset of the bar-mode instability to focus on the long-term
  behavior of each configuration.}
  \label{fig:evo_fluid_convergence}
\end{figure}

\subsection{Convergence of Dynamical Evolutions --- Polytropic EOS}

In this section, we present a convergence analysis of the dynamical
evolution of the \sbsix and \uteen configurations using the \etk. To do
so, we first consider initial data solutions using the \fuka
XCTS solver, with resolution $R = 17$.
The initial data is then evolved using the evolution setup described in
Sec.~\ref{sec:methods}, where the lowest spatial resolution used has a finest grid
spacing of $\Delta x = 0.4$, and the highest resolution has a finest grid
spacing of $\Delta x = 0.32$. In Fig.~\ref{fig:evo_H_convergence}, we
illustrate the Hamiltonian constraint violations along the $x$-axis at $t
= 41 M$ for the \sbsix (top) and \uteen (bottom) configurations.
We find the constraint violations to converge towards zero at second order
(green) which is expected given the algorithms used within \igm.

In Fig.~\ref{fig:evo_fluid_convergence} we extend our analysis to further
assess the impact of the initial data on the long-term dynamical
evolution. Here we include a convergence analysis of the $m = 2$ Fourier
mode, a key diagnostic for assessing the growth of the bar-mode
instability, for the \uteen configuration (for \sbsix see
Fig.~\ref{fig:evo_fluid_convergence_sb6} in Appendix~\ref{sec:appendix})
with XCTS (left panels) and QIC (right panels) solutions, at
varying initial data resolutions of $R = \left\lbrace 11 \,, 13 \,, 17
\right\rbrace$.

In the bottom panels we compute the relative difference in the power of
the $m = 2$ mode between high resolution simulations ($\Delta_H x =
0.32$) using initial data resolutions $R \in \left\lbrace 11 \,,
13\right\rbrace$ (referred to as \texttt{U13.11.H}, \texttt{U13.13.H}
respectively) (blue, orange lines respectively) to that of the highest ID
($R = 17$) and evolution resolution $\Delta_H x = 0.32$ (referred to as
\texttt{U13.17.H}). To determine the impact of the evolution resolution,
we compute the relative difference between the highest ID resolution
evolved using $\Delta_M x = 0.4$ (referred to as \texttt{U13.17.M}) to
that computed using $\Delta_H x = 0.32$ (\texttt{U13.17.H}).  For
clarity, we have aligned the datasets at the onset of the bar-mode
instability to focus on the long-term behavior of each configuration (see
Appendix~\ref{sec:appendix:longterm} for further discussion).

During the early evolution time of $t \lesssim 5$ms, we find that the
relative difference in the measured $m = 2$ Fourier modes resulting from
simulations using $R = 11$ and $R = 13$ initial data solutions against the simulation
using $R = 17$ to be essentially identical for both XCTS and QIC solutions. The
growth of these differences is consistent throughout the simulation with
only minor differences at later times. Using \texttt{U13.17.M}
(green line), we ascertain that the growth and saturation of the relative
difference at $\sim 10^0$ is dominated by the evolution resolution, and
the initial data resolution has minimal impact on the long-term
evolution of the $m = 2$ mode.

In the top panels of Fig.~\ref{fig:evo_fluid_convergence}, we illustrate
the evolution of the $m = 2$ Fourier mode for the \uteen configuration
for the XCTS (left) and QIC (right) solutions. Both are consistent with
behavior reported in Ref.~\cite{Baiotti2006} (see Fig.~7 therein) for the
duration of the simulation.

\begin{figure}[t]
  \includegraphics[width=\columnwidth]{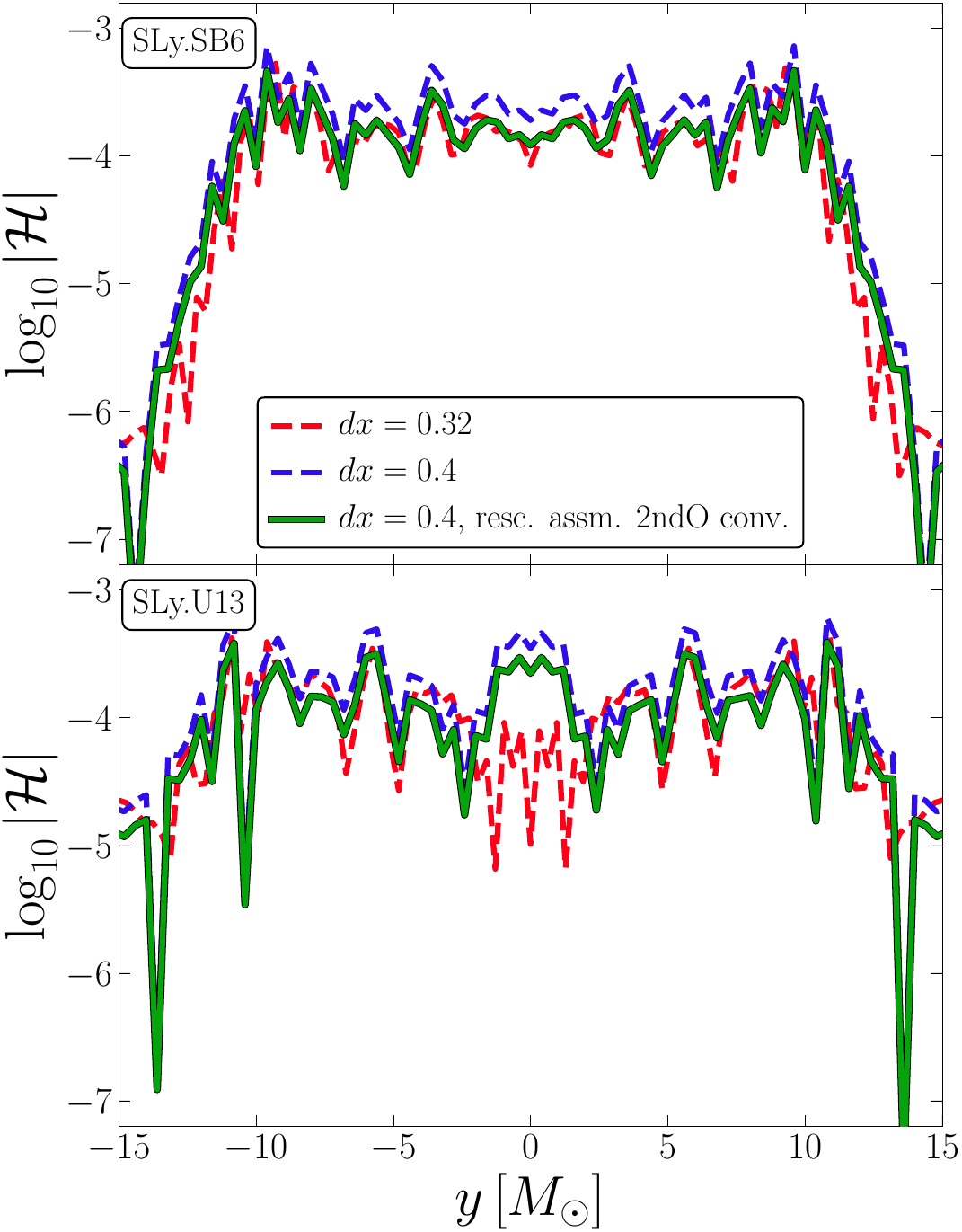}
  \caption{ Same as Fig~\ref{fig:evo_H_convergence} only here we analyze
  the \ddsbsix (\textit{top}) and \dduteen (\textit{bottom})
  configurations using the SLy EOS.}
  \label{fig:evo_H_convergence_SLy}
\end{figure}

\subsection{Convergence of Dynamical Evolutions --- Tabulated EOS}
As a final demonstration of the capabilities of \fuka to compute initial
data using a finite-temperature, tabulated equation of state, we present
a convergence analysis of the dynamical evolution of the \dduteen
configuration using the \etk. Here we restrict our analysis to ID
produced using the XCTS solver, the SLy EOS and the same evolution setup
as described in Sec.~\ref{sec:methods}.

In Fig.~\ref{fig:evo_H_convergence_SLy}, we illustrate the Hamiltonian
constraint violations along the $x$-axis at $t = 41 M$ for the \ddsbsix
(top) and \dduteen (bottom) configurations. Overall we recover the
expected second-order convergence (green) as predicted by the
hydrodynamics evolution, thought with some discrepancies at the stellar
center for the \uteen configuration.  Additionally, we observe an overall
higher magnitude of the constraint violations in the stellar interior
compared to the polytropic EOS solutions. We attribute the higher
constraint violations of the initial data slice (as measured in the
evolution gauge) for the tabulated EOS configurations as being primarily
due to the finite resolution of the SLy EOS and more realistic features
than a polytropic EOS.

\begin{figure}[t]
  \includegraphics[width=\columnwidth]{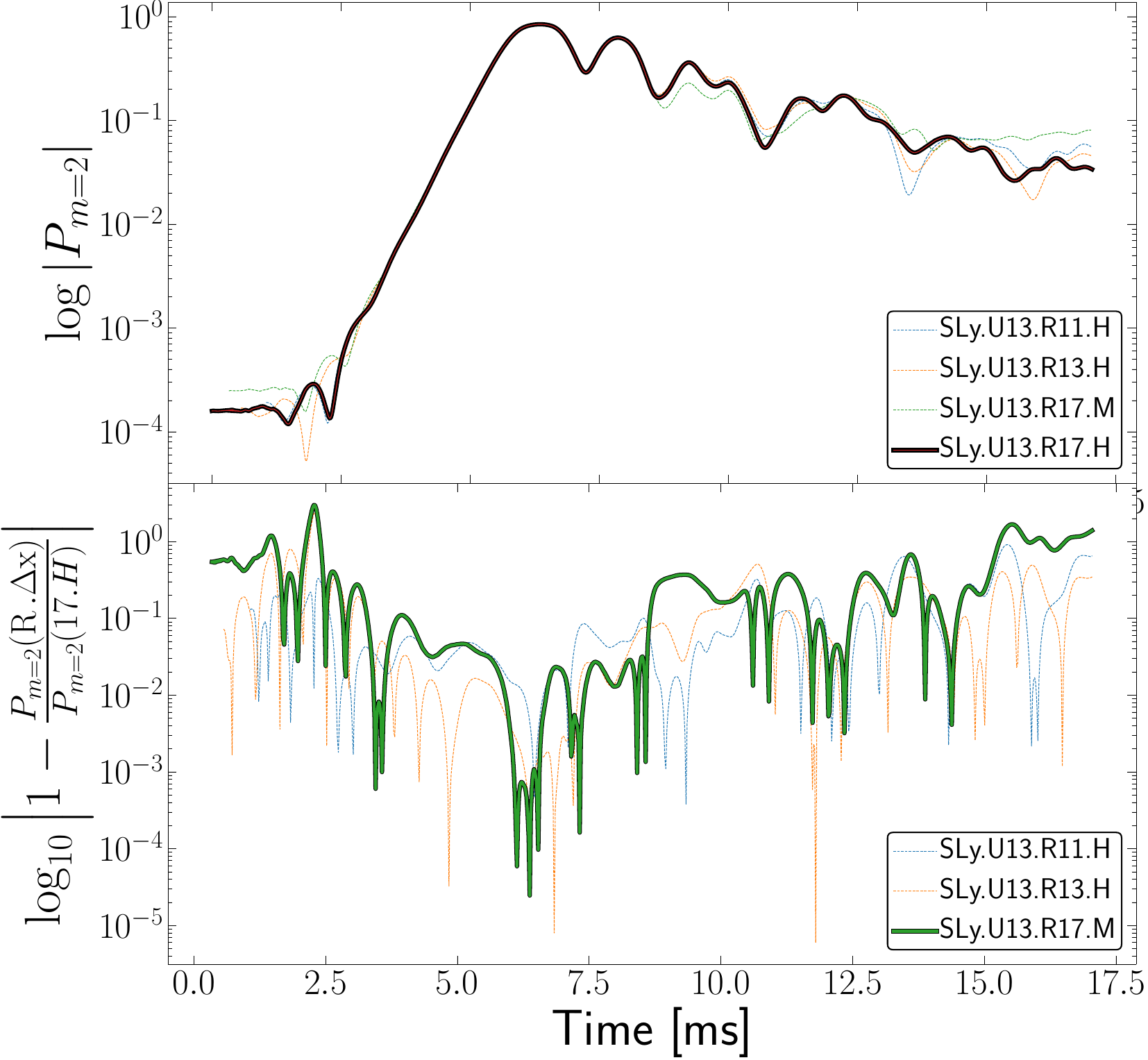}
  \caption{ Same as Fig~\ref{fig:evo_fluid_convergence} only here we
  focus on the \dduteen configuration and the data is again aligned at
  the onset of the bar-mode instability.}
  \label{fig:evo_fluid_convergence_SLy}
\end{figure}

To ascertain the impact of the initial data on the long-term dynamical
evolution of the \dduteen configuration, we present a convergence
analysis of the $m = 2$ Fourier mode in
Fig.~\ref{fig:evo_fluid_convergence_SLy}. Compared to
Fig.~\ref{fig:evo_fluid_convergence}, we see a similar trend in the
relative differences as a function of ID resolution (blue, orange lines)
and evolution resolution (green line). Here we again find that the
evolution resolution dominates the error growth, however, we find that
the initial data resolution has a slightly more pronounced impact at the
start of the simulation compared to the polytropic EOS solutions.  We
attribute this to the finite resolution and more rich features of the
tabulated EOS as compared to the simply polytropic EOS. Thus, we observe
a differences in the relaxation of the initial data slice
early in the evolution, but this quickly settles and the long-term
evolution is then dominated by the evolution resolution.

\section{Conclusion}
\label{sec:conclusion}
Precise and reliable initial data capable of accurately modeling
astrophysically relevant systems is an essential component to enable
accurate numerical relativity simulations with predictive power. To
facilitate the study of differentially rotating neutron stars, we
presented the recent extension of the \fuka public suite of initial data
codes using two different formulations of Einstein's field equations: the
QIC and XCTS systems of equations. Both solvers leverage the accuracy of
spectral methods using the \kadath library as well as \fuka's robust
framework for reliably computing initial data solutions. Furthermore,
\fuka includes a modular EOS framework enabling the use of piecewise
polytropic EOSs, tabulated EOSs in the \texttt{LORENE} format and, in the
latest version of \fuka, 3D tabulated EOSs in the \textit{stellar
collapse} format.

To validate our solutions we demonstrated the self-convergence of
each solver independently to ensure that we obtain the expected spectral
convergence properties of the numerical methods. We found that
spectral convergence is achieved for both solvers up to $\tilde{N} \sim
22$ at which point the solution deteriorates due to spectral noise, a
common issue with modeling steep gradients and discontinuities with
spectral methods. Additionally, we compared the solutions obtained
from both solvers and find excellent agreement despite considerable
difference in computational and algorithmic complexity. Finally, we
compared the QIC and XCTS solutions for the \sbsix and \uteen
configurations against previously published results and find excellent
agreement to three significant digits.

With confidence in the solutions computed with \fuka, we compared
the QIC solutions against those computed using the well-established \rns
code for a variety of differential rotation profiles and axis ratios.
Overall, we find excellent agreement between the two codes, with
deviations becoming pronounced for configurations on the unstable branch
of the solution space. Additionally, we note that for extreme
configurations ($\hat{\mathcal{A}} = 2.0$ and $r_p / r_e = 0.4$), we were
unable to compute solutions using \fuka. However, we leave this as an
open question for future study, given the stability of solutions in this
region of the parameter space, and the inconsistent trend we observed in \rns.

To explore the accuracy and impact of the computed ID solutions on the
long-term dynamical evolution, we conducted a convergence analysis
of the \sbsix and \uteen configurations using the \etk. Here we
demonstrate the expected second-order convergence of the Hamiltonian
constraint violations as well as the convergence of the $m = 2$ Fourier
mode, a key diagnostic for assessing the growth
of the bar-mode instability. We found that the initial data resolution has
a minimal impact on the long-term evolution of the $m = 2$ mode, with the
evolution resolution dominating the error growth and amplitude. Finally,
we found that the dynamical simulations qualitatively agree with
previously published results~\cite{Baiotti2006, Manca2007} (see
Fig.~\ref{fig:evo_fluid_convergence}).

To demonstrate the ability of \fuka to compute solutions using a
finite-temperature, tabulated equation of state, we computed initial
data solutions using the SLy EOS and conducted a similar convergence
analysis of the dynamical evolution. While we find higher constraint
violations compared to the polytropic EOS solutions, we still recover the
expected second-order convergence of the Hamiltonian constraint
violations. We attribute the higher constraint violations of the initial
data slice (as measured in the evolution gauge) for the tabulated EOS
configurations primarily due to the finite resolution and rich
features of the SLy EOS.

In this work, we demonstrated the accuracy and robustness for computing
differentially rotating neutron star initial data using \fuka, however,
there are still many avenues for future work. First, it would be
essential to extend \fuka to support additional rotation profiles such as
the extended KEH law~\cite{Uryu2017}, Uryu 8/9 laws~\cite{Uryu2017}, and
the more recent CR laws~\cite{Cassing2024}. Additionally, it would be
beneficial to extend the current implementation of the XCTS solver to
self-consistently support magnetic fields, which has been explored in the
public \texttt{LORENE}~\cite{Bocquet1995, Novak2003} and
\texttt{XNS}~\cite{DelZanna2007} frameworks as well as the closed source
\texttt{COCAL} code~\cite{Uryu2019}. Such extensions would serve to
bolster the findings in existing literature as well as enable exploration
across a broader range of EOSs, rotation profiles, and fluid profiles
(e.g., magnetic fields, temperature gradients). Finally, the existing
codes already provide an efficient means to explore the bar-mode
instability in differentially rotating neutron stars using tabulated
equations of state. Such studies would provide novel insights into the
impact of realistic microphysics on the growth and saturation of the
bar-mode instability as well as the resulting phenomenological
signatures.

\section*{Acknowledgements}
ST gratefully acknowledges support from NASA award ATP-80NSSC22K1898 and
support from the University of Idaho P3-R1 Initiative. TPJ acknowledges support from
NASA FINESST-80NSSC23K1437. MC acknowledges
support from the European Research Council Advanced Grant ‘JETSET: Launching,
propagation and emission of relativistic jets from binary mergers and across
mass scales’ (grant no. 884631).
This research made use of
Idaho National Laboratory's High Performance Computing systems located at
the Collaborative Computing Center and supported by the Office of Nuclear
Energy of the U.S. Department of Energy and the Nuclear Science User
Facilities under Contract No. DE-AC07-05ID14517. Finally, this work
benefited from the extensive use of the open-source packages
NumPy~\cite{NumPy}, Matplotlib~\cite{Matplotlib}, and Kuibit~\cite{Kuibit}.
\newpage
\begin{appendix}
\section{Polytropic \sbsix Analysis}
\label{sec:appendix}
In this appendix, we include the convergence analysis of the $m = 2$
Fourier mode during dynamical evolution for the
\sbsix configuration to assess the impact of initial data resolution $R =
\left\lbrace11 \,, 13 \,, 17\right\rbrace$ with evolution resolution
$\Delta_H x = 0.32$ and $\Delta_M x = 0.4$.

\begin{figure}[t]
  \includegraphics[width=\columnwidth]{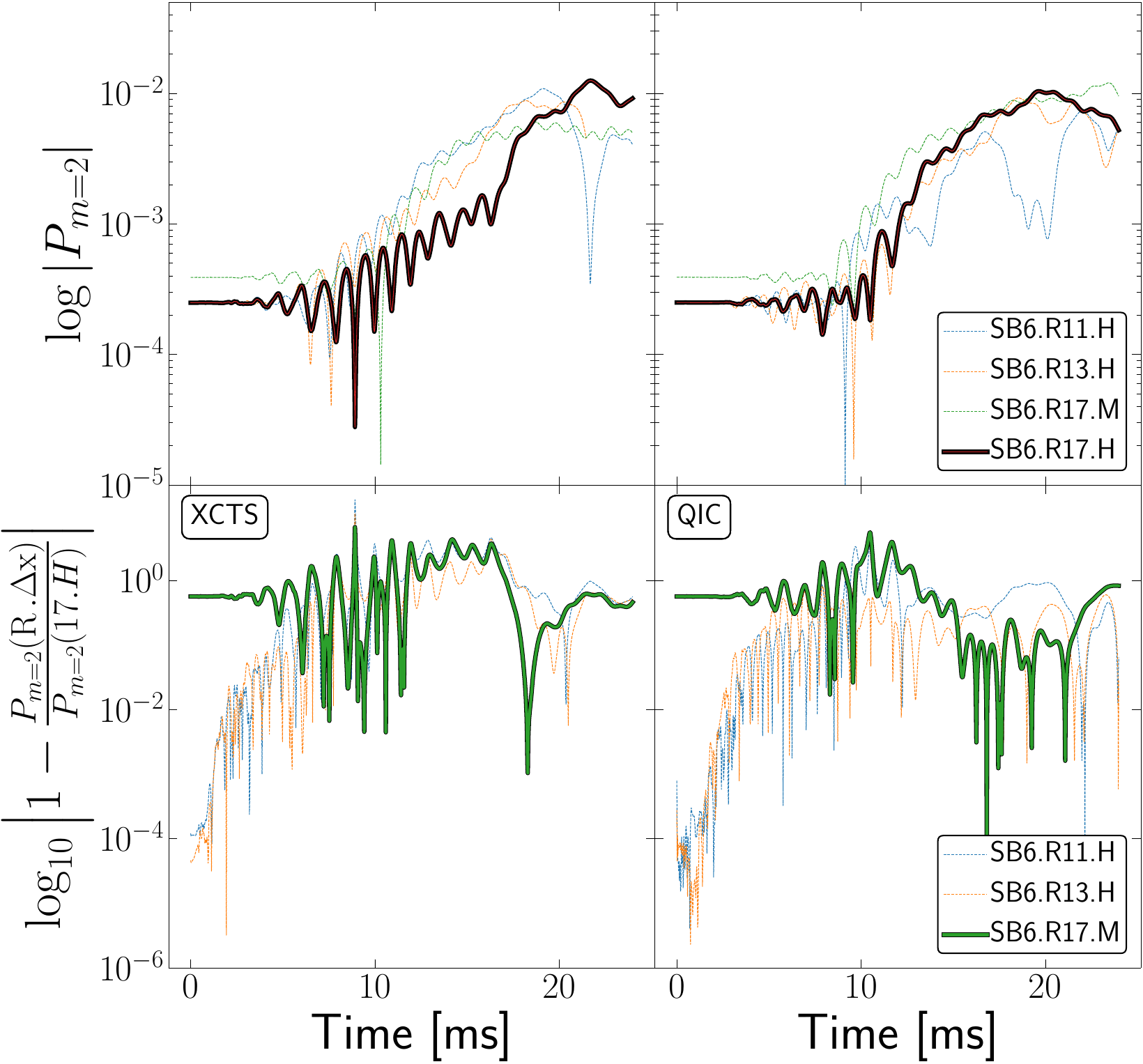}
  \caption{ Same as Fig.~\ref{fig:evo_fluid_convergence}, but for the \sbsix
  configuration.}
  \label{fig:evo_fluid_convergence_sb6}
\end{figure}

As shown in Fig.~\ref{fig:evo_fluid_convergence_sb6}, we find similar
behavior as that observed for the \uteen configuration in
Fig.~\ref{fig:evo_fluid_convergence}. Specifically, during the early
evolution time of $t \lesssim 8$ms, we find that the relative difference
in the measured $m = 2$ Fourier modes resulting from simulations using $R
= 11$ and $R = 13$ initial data solutions against the simulation using $R = 17$
to have a minor influence on the resulting dynamics. Unlike \uteen, the \sbsix
configuration is stable to the bar-mode instability, and thus is more
sensitive to small perturbations. Therefore, we do find larger
discrepancies between the very coarse $R = 11$ initial data solution
compared to the higher resolution solutions. This trend is consistent
between QIC and XCTS solutions, further validating this hypothesis.

\section{Discussion on Data Alignment}
\label{sec:appendix:longterm}

In Sec.~\ref{sec:results}, we presented a convergence analysis of the $m
= 2$ Fourier mode during dynamical evolution for
the \uteen and \dduteen configurations (see
Figs.~\ref{fig:evo_fluid_convergence} and
\ref{fig:evo_fluid_convergence_SLy}, respectively). In both cases, we
aligned the datasets at the onset of the bar-mode instability to focus on
the long-term behavior of the each configuration and ascertain the ID
impact on the long-term behavior of the system.

\begin{figure}[t]
  \includegraphics[width=\columnwidth]{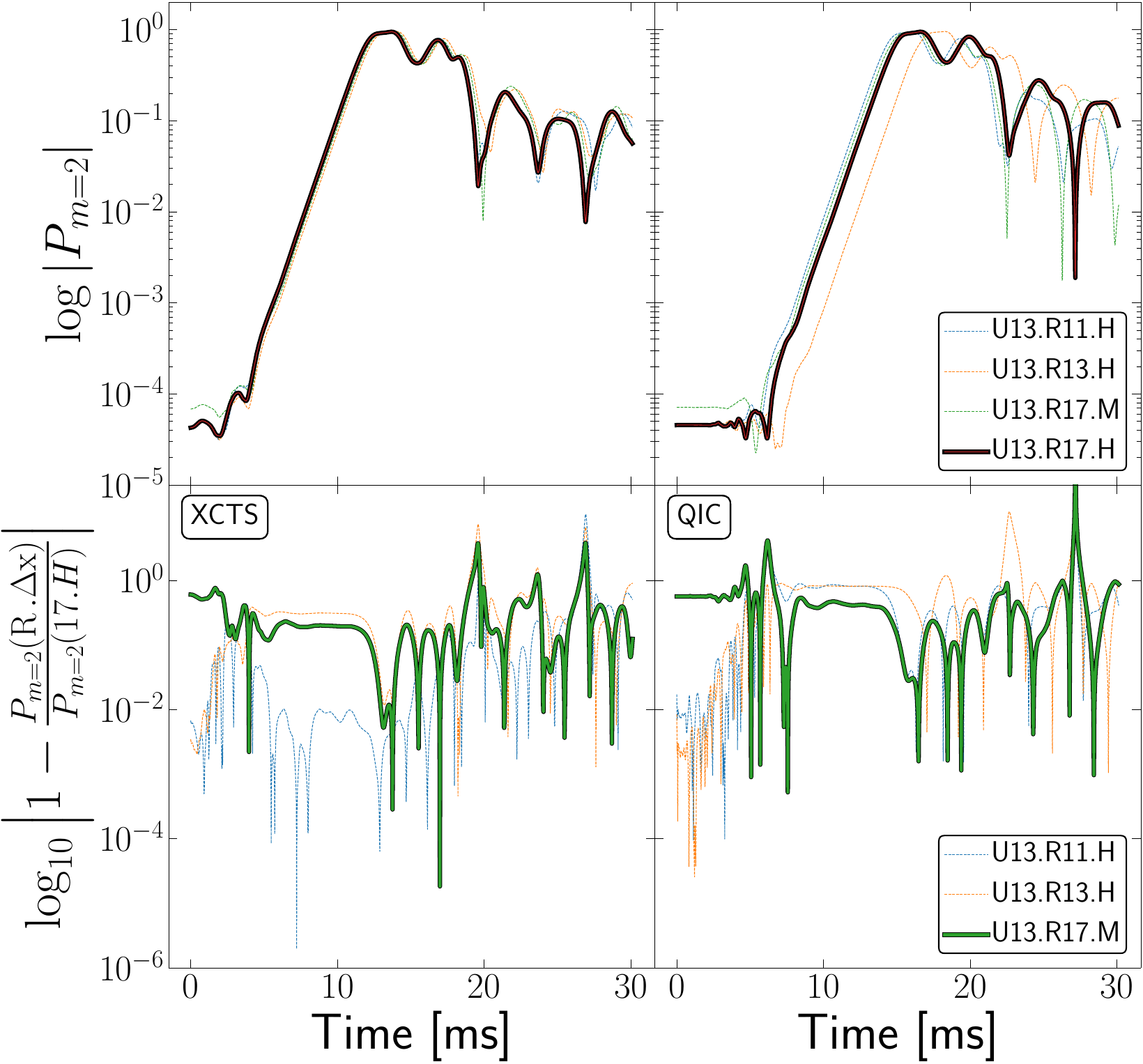}
  \caption{ Same as Fig~\ref{fig:evo_fluid_convergence} only the \uteen
  data is not aligned.}
  \label{fig:evo_fluid_convergence_noshift}
\end{figure}

\begin{figure}[t]
  \includegraphics[width=\columnwidth]{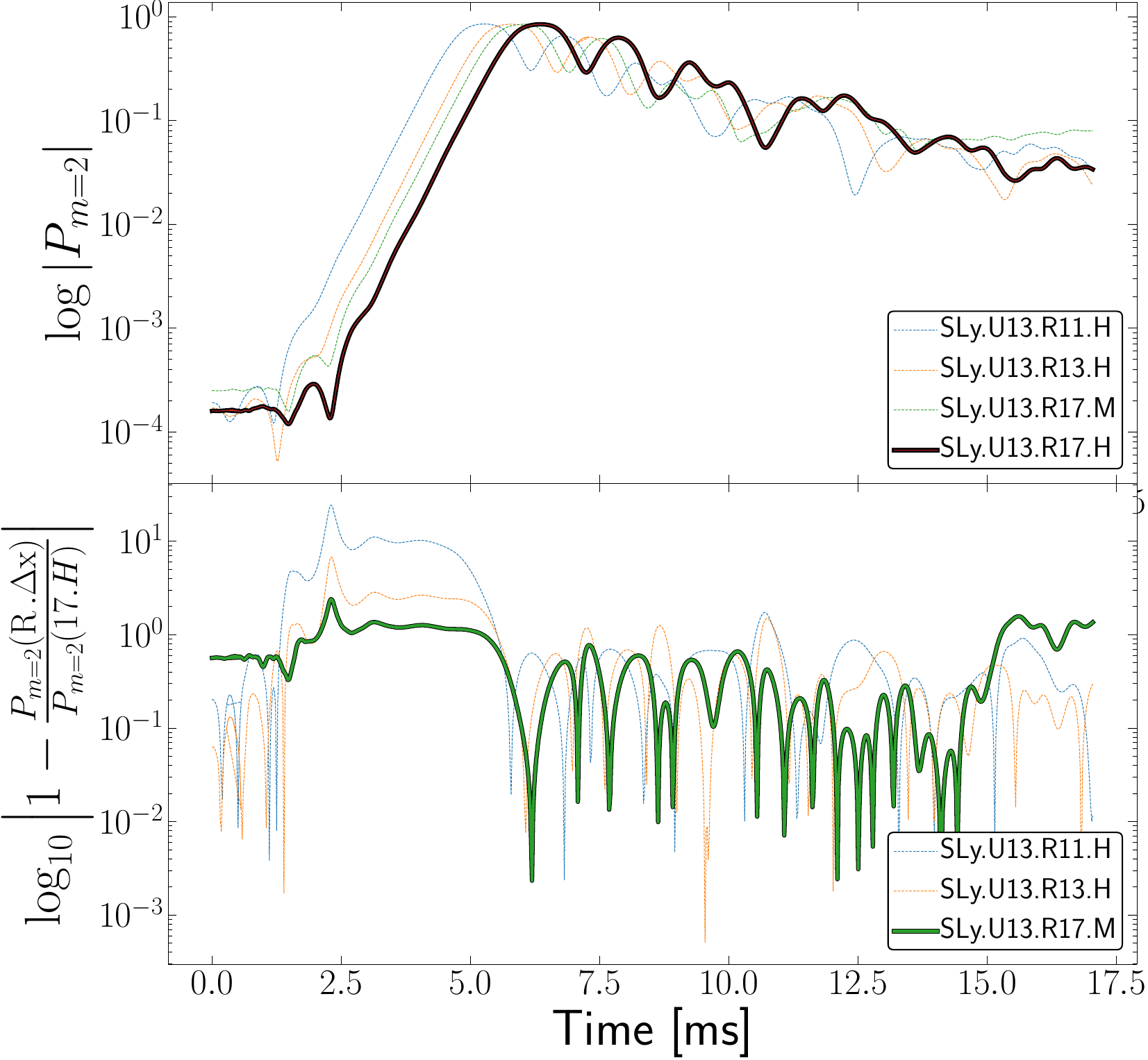}
  \caption{ Same as Fig~\ref{fig:evo_fluid_convergence_SLy} only the data
  is not aligned.}
  \label{fig:evo_fluid_convergence_SLy_noshift}
\end{figure}

To elaborate on this choice, we provide in
Figs.~\ref{fig:evo_fluid_convergence_noshift} and
\ref{fig:evo_fluid_convergence_SLy_noshift} the same data presented in
Figs.~\ref{fig:evo_fluid_convergence} and
\ref{fig:evo_fluid_convergence_SLy}, respectively, only now the datasets
are not aligned at the onset of the bar-mode instability. In the case of
Fig.~\ref{fig:evo_fluid_convergence_noshift}, we find that the
transformation of the QIC solution from Spherical to Cartesian
coordinates leads to slightly different onset time for the bar-mode
instability due to higher constraint violations near coordinate
singularities. As these configurations are only marginally stable,
differences in the initial perturbations lead to variations in the onset
time of the instability, however, the long-term behavior is consistent
for all ID resolutions.

Conversely, in the case of
Fig.~\ref{fig:evo_fluid_convergence_SLy_noshift}, where only the XCTS
solution is considered, we find that the differences in the initial data
resolution can have a pronounced impact on the onset time of the bar-mode
instability. Here, the differences in the initial perturbations are
largely attributed to the finite resolution and more rich features of the
tabulated EOS as compared to the simply polytropic EOS.

With these observations in mind, we chose to align the datasets at the
onset of the bar-mode instability in order to focus on the long-term
impact the ID resolution has on the system rather than on the initial
perturbations that seed the instability.

\end{appendix}

\bibliographystyle{apsrev4-2}
\bibliography{refs}

@article{Abbott2017b,
   author = {{The LIGO Scientific Collaboration} and {the Virgo Collaboration} and {Abbott}, B.~P. and {Abbott}, R. and {Abbott}, T.~D. and {Acernese}, F. and {Ackley}, K. and a{Adams}, C. and {Adams}, T. and {Addesso}, P. and
         {Adhikari}, R.~X. and {Adya}, V.~B. and et al.},
  collaboration = {LIGO Scientific Collaboration and Virgo Collaboration},
  title={Multi-messenger Observations of a Binary Neutron Star Merger},
  journal={Astrophys. J. Lett.},
         year = "2017",
        month = oct,
       volume = {848},
       number = {2},
          eid = {L12},
        pages = {L12},
          doi = {10.3847/2041-8213/aa91c9},
archivePrefix = {arXiv},
       eprint = {1710.05833},
 primaryClass = {astro-ph.HE},
       adsurl = {https://ui.adsabs.harvard.edu/abs/2017ApJ...848L..12A}
}

@article{Abdikamalov2009aq,
	Adsurl = {http://adsabs.harvard.edu/abs/2010PhRvD..81d4012A},
	Archiveprefix = {arXiv},
	Author = {{Abdikamalov}, E.~B. and {Ott}, C.~D. and {Rezzolla}, L. and {Dessart}, L. and {Dimmelmeier}, H. and {Marek}, A. and {Janka}, H.-T.},
	Doi = {10.1103/PhysRevD.81.044012},
	Eid = {044012},
	Eprint = {0910.2703},
	Journal = {Phys. Rev. D},
	Month = feb,
	Number = 4,
	Pages = {044012},
	Primaryclass = {astro-ph.HE},
	Title = {{Axisymmetric general relativistic simulations of the accretion-induced collapse of white dwarfs}},
	Volume = 81,
	Year = 2010}

@article{Alcubierre:2002kk,
    author = "Alcubierre, Miguel and Bruegmann, Bernd and Diener, Peter and Koppitz, Michael and Pollney, Denis and Seidel, Edward and Takahashi, Ryoji",
    title = "{Gauge conditions for long term numerical black hole evolutions without excision}",
    eprint = "gr-qc/0206072",
    archivePrefix = "arXiv",
    reportNumber = "AEI-2002-037",
    doi = "10.1103/PhysRevD.67.084023",
    journal = "Phys. Rev. D",
    volume = "67",
    pages = "084023",
    year = "2003"
}

@article{Ansorg2009,
	Adsurl = {http://adsabs.harvard.edu/abs/2009MNRAS.396.2359A},
	Archiveprefix = {arXiv},
	Author = {{Ansorg}, M. and {Gondek-Rosi{\'n}ska}, D. and {Villain}, L.},
	Doi = {10.1111/j.1365-2966.2009.14904.x},
	Eprint = {0812.3347},
	Journal = {Mon. Not. R. Astron. Soc.},
	Month = jul,
	Pages = {2359-2366},
	Primaryclass = {gr-qc},
	Title = {{On the solution space of differentially rotating neutron stars in general relativity}},
	Volume = 396,
	Year = 2009}

@article{Baiotti2016,
      author         = "Baiotti, Luca and Rezzolla, Luciano",
      title          = "{Binary neutron-star mergers: a review of Einstein's
                        richest laboratory}",
      journal        = "Rept. Prog. Phys.",
      volume         = "80",
      year           = "2017",
      number         = "9",
      pages          = "096901",
      doi            = "10.1088/1361-6633/aa67bb",
      eprint         = "1607.03540",
      archivePrefix  = "arXiv",
      primaryClass   = "gr-qc",
      SLACcitation   = "%%CITATION = ARXIV:1607.03540;%%"
}

@article{Baker:2005vv,
    author = "Baker, John G. and Centrella, Joan and Choi, Dae-Il and Koppitz, Michael and van Meter, James",
    title = "{Gravitational wave extraction from an inspiraling configuration of merging black holes}",
    eprint = "gr-qc/0511103",
    archivePrefix = "arXiv",
    doi = "10.1103/PhysRevLett.96.111102",
    journal = "Phys. Rev. Lett.",
    volume = "96",
    pages = "111102",
    year = "2006"
}

@article{Bamber:2024qzi,
    author = "Bamber, Jamie and Tsokaros, Antonios and Ruiz, Milton and Shapiro, Stuart L.",
    title = "{Postmerger multimessenger analysis of binary neutron stars: Effect of the magnetic field strength and topology}",
    eprint = "2411.00943",
    archivePrefix = "arXiv",
    primaryClass = "gr-qc",
    doi = "10.1103/PhysRevD.111.044038",
    journal = "Phys. Rev. D",
    volume = "111",
    number = "4",
    pages = "044038",
    year = "2025"
}

@article{Baumgarte1998,
    author = "Baumgarte, Thomas W. and Shapiro, Stuart L.",
    title = "{On the numerical integration of Einstein's field equations}",
    eprint = "gr-qc/9810065",
    archivePrefix = "arXiv",
    doi = "10.1103/PhysRevD.59.024007",
    journal = "Phys. Rev. D",
    volume = "59",
    pages = "024007",
    year = "1998"
}

@article{Baumgarte00b,
   author = {{Baumgarte}, T.~W. and {Shapiro}, S.~L. and {Shibata}, M.},
    title = "{On the Maximum Mass of Differentially Rotating Neutron Stars}",
  journal = {Astrophys. J. Lett.},
   eprint = {astro-ph/9910565},
     year = 2000,
    month = jan,
   volume = 528,
    pages = {L29-L32},
      doi = {10.1086/312425},
   adsurl = {http://adsabs.harvard.edu/abs/2000ApJ...528L..29B}
}

@book{Baumgarte2010,
  address =       {Cambridge, UK},
  author =        {{Baumgarte}, T.~W. and {Shapiro}, S.~L.},
  booktitle =     {Numerical Relativity: Solving Einstein's Equations on
                   the Computer},
  publisher =     {Cambridge University Press},
  title =         {{Numerical Relativity: Solving Einstein's Equations
                   on the Computer}},
  year =          {2010},
  doi =           {10.1017/cbo9781139193344},
}

@article{Baiotti2006,
    author = "Baiotti, Luca and De Pietri, Roberto and Manca, Gian Mario and Rezzolla, Luciano",
    title = "{Accurate simulations of the dynamical barmode instability in full General Relativity}",
    eprint = "astro-ph/0609473",
    archivePrefix = "arXiv",
    doi = "10.1103/PhysRevD.75.044023",
    journal = "Phys. Rev. D",
    volume = "75",
    pages = "044023",
    year = "2007"
}

@article{Baiotti08,
  Adsurl = {http://adsabs.harvard.edu/abs/2008PhRvD..78h4033B},
  Archiveprefix = {arXiv},
  Author = {{Baiotti}, L. and {Giacomazzo}, B. and {Rezzolla}, L.},
  Doi = {10.1103/PhysRevD.78.084033},
  Eprint = {0804.0594},
  Journal = {Phys. Rev. D},
  Month = oct,
  Number = 8,
  Pages = {084033},
  Primaryclass = {gr-qc},
  Title = {{Accurate evolutions of inspiralling neutron-star binaries: Prompt and delayed collapse to a black hole}},
  Volume = 78,
  Year = 2008
}

@article{Bocquet1995,
    author = "Bocquet, M. and Bonazzola, S. and Gourgoulhon, E. and Novak, J.",
    title = "{Rotating neutron star models with magnetic field}",
    eprint = "gr-qc/9503044",
    archivePrefix = "arXiv",
    journal = "Astron. Astrophys.",
    volume = "301",
    pages = "757",
    year = "1995"
}

@ARTICLE{Bozzola2017,
       author = {{Bozzola}, Gabriele and {Stergioulas}, Nikolaos and {Bauswein}, Andreas},
        title = "{Universal relations for differentially rotating relativistic stars at the threshold to collapse}",
      journal = {Mon. Not. R. Astron. Soc.},
         year = 2018,
        month = mar,
       volume = {474},
       number = {3},
        pages = {3557-3564},
          doi = {10.1093/mnras/stx3002},
archivePrefix = {arXiv},
       eprint = {1709.02787},
 primaryClass = {gr-qc},
       adsurl = {https://ui.adsabs.harvard.edu/abs/2018MNRAS.474.3557B}
}

@ARTICLE{Bozzola2019,
       author = {{Bozzola}, Gabriele and {Espino}, Pedro L. and {Lewin}, Collin D. and {Paschalidis}, Vasileios},
        title = "{Maximum mass and universal relations of rotating relativistic hybrid hadron-quark stars}",
      journal = {European Physical Journal A},
         year = 2019,
        month = sep,
       volume = {55},
       number = {9},
          eid = {149},
        pages = {149},
          doi = {10.1140/epja/i2019-12831-2},
archivePrefix = {arXiv},
       eprint = {1905.00028},
 primaryClass = {astro-ph.HE},
       adsurl = {https://ui.adsabs.harvard.edu/abs/2019EPJA...55..149B}
}

@article{Breu2016,
	Adsurl = {http://adsabs.harvard.edu/abs/2016MNRAS.459..646B},
	Archiveprefix = {arXiv},
	Author = {{Breu}, C. and {Rezzolla}, L.},
	Doi = {10.1093/mnras/stw575},
	Eprint = {1601.06083},
	Journal = {Mon. Not. R. Astron. Soc.},
	Month = jun,
	Pages = {646-656},
	Primaryclass = {gr-qc},
	Title = {{Maximum mass, moment of inertia and compactness of relativistic stars}},
	Volume = 459,
	Year = 2016}

@article{Burrows07a,
	Adsurl = {http://adsabs.harvard.edu/abs/2007ApJ...664..416B},
	Author = {{Burrows}, A. and {Dessart}, L. and {Livne}, E. and {Ott}, C.~D. and {Murphy}, J.},
	Doi = {10.1086/519161},
	Eprint = {arXiv:astro-ph/0702539},
	Journal = {Astrophys. J.},
	Month = jul,
	Pages = {416-434},
	Title = {{Simulations of Magnetically Driven Supernova and Hypernova Explosions in the Context of Rapid Rotation}},
	Volume = 664,
	Year = 2007}

@article{Camelio2021,
  title = {Axisymmetric models for neutron star merger remnants with realistic thermal and rotational profiles},
  author = {Camelio, Giovanni and Dietrich, Tim and Rosswog, Stephan and Haskell, Brynmor},
  journal = {Phys. Rev. D},
  volume = {103},
  issue = {6},
  pages = {063014},
  numpages = {17},
  year = {2021},
  month = {Mar},
  publisher = {American Physical Society},
  doi = {10.1103/PhysRevD.103.063014},
  url = {https://link.aps.org/doi/10.1103/PhysRevD.103.063014}
}

@article{Campanelli2005,
    author = "Campanelli, Manuela and Lousto, C. O. and Marronetti, P. and Zlochower, Y.",
    title = "{Accurate evolutions of orbiting black-hole binaries without excision}",
    eprint = "gr-qc/0511048",
    archivePrefix = "arXiv",
    doi = "10.1103/PhysRevLett.96.111101",
    journal = "Phys. Rev. Lett.",
    volume = "96",
    pages = "111101",
    year = "2006"
}

@ARTICLE{Cassing2024,
       author = {{Cassing}, Marie and {Rezzolla}, Luciano},
        title = "{Realistic models of general-relativistic differentially rotating stars}",
      journal = {Mon. Not. R. Astron. Soc.},
         year = 2024,
        month = jul,
       volume = {532},
       number = {1},
        pages = {945-964},
          doi = {10.1093/mnras/stae1527},
archivePrefix = {arXiv},
       eprint = {2405.06609},
 primaryClass = {gr-qc},
       adsurl = {https://ui.adsabs.harvard.edu/abs/2024MNRAS.532..945C}
}

@article{Chabanat1997,
    author = "Chabanat, E. and Bonche, P. and Haensel, P. and Meyer, J. and Schaeffer, R.",
    title = "{A Skyrme parametrization from subnuclear to neutron star densities. 2. Nuclei far from stablities}",
    doi = "10.1016/S0375-9474(98)00180-8",
    journal = "Nucl. Phys. A",
    volume = "635",
    pages = "231--256",
    year = "1998",
    note = "[Erratum: Nucl.Phys.A 643, 441--441 (1998)]"
}

@article{Chawhan:2025esv,
    author = "Chawhan, Pavan and Duez, Matthew D. and Foucart, Francois and Cheong, Patrick Chi-Kit and Muhammed, Nishad",
    title = "{Axisymmetric hydrodynamics in numerical relativity: treating coordinate singularity, artificial heating and modeling MHD instabilities}",
    eprint = "2510.14127",
    archivePrefix = "arXiv",
    primaryClass = "astro-ph.HE",
    month = "10",
    year = "2025",
    journal = "arXiv"
}

@article{Cheong2024,
    author = "Cheong, Patrick Chi-Kit and Muhammed, Nishad and Chawhan, Pavan and Duez, Matthew D. and Foucart, Francois and Kidder, Lawrence E. and Pfeiffer, Harald P. and Scheel, Mark A.",
    title = "{High angular momentum hot differentially rotating equilibrium star evolutions in conformally flat spacetime}",
    eprint = "2402.18529",
    archivePrefix = "arXiv",
    primaryClass = "astro-ph.HE",
    doi = "10.1103/PhysRevD.110.043015",
    journal = "Phys. Rev. D",
    volume = "110",
    number = "4",
    pages = "043015",
    year = "2024"
}

@article{Cook92b,
       author = {{Cook}, Gregory B. and {Shapiro}, Stuart L. and {Teukolsky}, Saul A.},
        title = "{Spin-up of a Rapidly Rotating Star by Angular Momentum Loss: Effects of General Relativity}",
      journal = {Astrophys. J.},
         year = "1992",
        month = oct,
       volume = {398},
        pages = {203},
          doi = {10.1086/171849},
       adsurl = {https://ui.adsabs.harvard.edu/abs/1992ApJ...398..203C}
}

@article{Corman2024,
    author = "Corman, Maxence and East, William E.",
    title = "{Black hole-neutron star mergers in Einstein-scalar-Gauss-Bonnet gravity}",
    eprint = "2405.18496",
    archivePrefix = "arXiv",
    primaryClass = "gr-qc",
    doi = "10.1103/PhysRevD.110.084065",
    journal = "Phys. Rev. D",
    volume = "110",
    number = "8",
    pages = "084065",
    year = "2024"
}

@article{DelZanna2007,
    author = "Del Zanna, L. and Zanotti, O. and Bucciantini, N. and Londrillo, P.",
    title = "{ECHO: an Eulerian Conservative High Order scheme for general relativistic magnetohydrodynamics and magnetodynamics}",
    eprint = "0704.3206",
    archivePrefix = "arXiv",
    primaryClass = "astro-ph",
    doi = "10.1051/0004-6361:20077093",
    journal = "Astron. Astrophys.",
    volume = "473",
    pages = "11--30",
    year = "2007"
}

@ARTICLE{Cupp2025,
       author = {{Cupp}, Samuel and {Werneck}, Leonardo R. and {Jacques}, Terrence Pierre and {Tootle}, Samuel and {Etienne}, Zachariah B.},
        title = "{GRHayL: a modern, infrastructure-agnostic, extensible library for GRMHD simulations}",
      journal = {arXiv e-prints},
         year = 2025,
        month = dec,
          eid = {arXiv:2512.15846},
        pages = {arXiv:2512.15846},
          doi = {10.48550/arXiv.2512.15846},
archivePrefix = {arXiv},
       eprint = {2512.15846},
 primaryClass = {gr-qc},
       adsurl = {https://ui.adsabs.harvard.edu/abs/2025arXiv251215846C}
}

@article{Dessart2006,
       author = {{Dessart}, L. and {Burrows}, A. and {Ott}, C.~D. and {Livne}, E. and {Yoon}, S. -C. and {Langer}, N.},
        title = "{Multidimensional Simulations of the Accretion-induced Collapse of White Dwarfs to Neutron Stars}",
      journal = {Astrophys. J.},
         year = 2006,
        month = jun,
       volume = {644},
       number = {2},
        pages = {1063-1084},
          doi = {10.1086/503626},
archivePrefix = {arXiv},
       eprint = {astro-ph/0601603},
 primaryClass = {astro-ph},
       adsurl = {https://ui.adsabs.harvard.edu/abs/2006ApJ...644.1063D}
}

@article{Duez:2006qe,
	Archiveprefix = {arXiv},
	Author = {Duez, Matthew D. and Liu, Yuk Tung and Shapiro, Stuart L. and Shibata, Masaru and Stephens, Branson C.},
	Doi = {10.1103/PhysRevD.73.104015},
	Eprint = {astro-ph/0605331},
	Journal = {Phys. Rev. D},
	Pages = {104015},
	Slaccitation = {%%CITATION = ASTRO-PH/0605331;%%},
	Title = {{Evolution of magnetized, differentially rotating neutron stars: Simulations in full general relativity}},
	Volume = {73},
	Year = {2006}}

@article{Espino2019,
  title = {Revisiting the maximum mass of differentially rotating neutron stars in general relativity with realistic equations of state},
  author = {Espino, Pedro L. and Paschalidis, Vasileios},
  journal = {Phys. Rev. D},
  volume = {99},
  issue = {8},
  pages = {083017},
  numpages = {16},
  year = {2019},
  month = {Apr},
  publisher = {American Physical Society},
  doi = {10.1103/PhysRevD.99.083017},
  url = {https://link.aps.org/doi/10.1103/PhysRevD.99.083017}
}

@article{Espino2019b,
  title = {Dynamical stability of quasitoroidal differentially rotating neutron stars},
  author = {Espino, Pedro L. and Paschalidis, Vasileios and Baumgarte, Thomas W. and Shapiro, Stuart L.},
  journal = {Phys. Rev. D},
  volume = {100},
  issue = {4},
  pages = {043014},
  numpages = {19},
  year = {2019},
  month = {Aug},
  publisher = {American Physical Society},
  doi = {10.1103/PhysRevD.100.043014},
  url = {https://link.aps.org/doi/10.1103/PhysRevD.100.043014}
}

@ARTICLE{Etienne2015,
       author = {{Etienne}, Zachariah B. and {Paschalidis}, Vasileios and {Haas}, Roland and {M{\"o}sta}, Philipp and {Shapiro}, Stuart L.},
        title = "{IllinoisGRMHD: an open-source, user-friendly GRMHD code for dynamical spacetimes}",
      journal = {Classical and Quantum Gravity},
         year = 2015,
        month = sep,
       volume = {32},
       number = {17},
          eid = {175009},
        pages = {175009},
          doi = {10.1088/0264-9381/32/17/175009},
archivePrefix = {arXiv},
       eprint = {1501.07276},
 primaryClass = {astro-ph.HE},
       adsurl = {https://ui.adsabs.harvard.edu/abs/2015CQGra..32q5009E}
}

@article{Franceschetti2022,
    author = "Franceschetti, Kevin and Del Zanna, Luca and Soldateschi, Jacopo and Bucciantini, Niccol{\`o}",
    title = "{Numerical Equilibrium Configurations and Quadrupole Moments of Post-Merger Differentially Rotating Relativistic Stars}",
    doi = "10.3390/universe8030172",
    journal = "Universe",
    volume = "8",
    number = "3",
    pages = "172",
    year = "2022"
}

@article{Friedman88,
	Author = {J. L. Friedman and J. R. Ipser and R. D. Sorkin},
	Journal = {Astrophys. J.},
	Key = {Friedman88},
	Pages = {722-724},
	Title = {Turning-Point Method for Axisymmetric Stability of Rotating Relativistic Stars},
	Volume = {325},
	Year = {1988}}

@article{Giacomazzo2011,
	Adsurl = {http://adsabs.harvard.edu/abs/2011PhRvD..84b4022G},
	Archiveprefix = {arXiv},
	Author = {{Giacomazzo}, B. and {Rezzolla}, L. and {Stergioulas}, N.},
	Doi = {10.1103/PhysRevD.84.024022},
	Eid = {024022},
	Eprint = {1105.0122},
	Journal = {Phys. Rev. D},
	Month = jul,
	Number = 2,
	Pages = {024022},
	Primaryclass = {gr-qc},
	Title = {{Collapse of differentially rotating neutron stars and cosmic censorship}},
	Volume = 84,
	Year = 2011}

@ARTICLE{Gondek2016,
   author = {{Gondek-Rosi{\'n}ska}, D. and {Kowalska}, I. and {Villain}, L. and
	{Ansorg}, M. and {Kucaba}, M.},
    title = "{A New View on the Maximum Mass of Differentially Rotating Neutron Stars}",
  journal = {Astrophys. J.},
archivePrefix = "arXiv",
   eprint = {1609.02336},
 primaryClass = "astro-ph.HE",
     year = 2017,
    month = mar,
   volume = 837,
      eid = {58},
    pages = {58},
      doi = {10.3847/1538-4357/aa56c1},
   adsurl = {http://adsabs.harvard.edu/abs/2017ApJ...837...58G}
}

@article{Gourgoulhon2007,
  author =        {{Gourgoulhon}, Eric},
  journal =       {arXiv e-prints},
  month =         mar,
  pages =         {gr-qc/0703035},
  title =         {{3+1 Formalism and Bases of Numerical Relativity}},
  year =          {2007},
  doi =           {10.48550/arXiv.gr-qc/0703035},
  eid =           {gr-qc/0703035},
}

@inproceedings{Gourgoulhon2010,
    author = "Gourgoulhon, Eric",
    title = "{An Introduction to the theory of rotating relativistic stars}",
    booktitle = "{CompStar 2010: School and Workshop on Computational Tools for Compact Star Astrophysics}",
    eprint = "1003.5015",
    archivePrefix = "arXiv",
    primaryClass = "gr-qc",
    month = "3",
    year = "2010"
}

@article{Goussard1998,
	Adsurl = {http://adsabs.harvard.edu/abs/1998A%26A...330.1005G},
	Author = {{Goussard}, {J.-O.} and {Haensel}, P. and {Zdunik}, J.~L.},
	Eprint = {arXiv:astro-ph/9711347},
	Journal = {Astron. and Astrophys.},
	Month = feb,
	Pages = {1005-1016},
	Title = {{Rapid differential rotation of protoneutron stars and constraints on radio pulsars periods}},
	Volume = 330,
	Year = 1998}

@article{Grandclement2009,
    author = "Grandclement, Philippe",
    title = "{Kadath: A Spectral solver for theoretical physics}",
    eprint = "0909.1228",
    archivePrefix = "arXiv",
    primaryClass = "gr-qc",
    doi = "10.1016/j.jcp.2010.01.005",
    journal = "J. Comput. Phys.",
    volume = "229",
    pages = "3334--3357",
    year = "2010"
}

@article{Grandclement2014,
    author = "Grandclement, Philippe and Som\'e, Claire and Gourgoulhon, Eric",
    title = "{Models of rotating boson stars and geodesics around them: new type of orbits}",
    eprint = "1405.4837",
    archivePrefix = "arXiv",
    primaryClass = "gr-qc",
    doi = "10.1103/PhysRevD.90.024068",
    journal = "Phys. Rev. D",
    volume = "90",
    number = "2",
    pages = "024068",
    year = "2014"
}

@article{Hanauske2016,
   author = {{Hanauske}, M. and {Takami}, K. and {Bovard}, L. and
                  {Rezzolla}, L. and {Font}, J.~A. and {Galeazzi}, F. and
                  {St{\"o}cker}, H.},
    title = "{Rotational properties of hypermassive neutron stars from binary mergers}",
  journal = {Phys. Rev. D},
archivePrefix = "arXiv",
   eprint = {1611.07152},
 primaryClass = "gr-qc",
     year = 2017,
    month = aug,
   volume = 96,
   number = 4,
      eid = {043004},
    pages = {043004},
      doi = {10.1103/PhysRevD.96.043004},
   adsurl = {http://adsabs.harvard.edu/abs/2017PhRvD..96d3004H}
}

@article{Jacques2024,
    author = "Jacques, Terrence Pierre and Cupp, Samuel and Werneck, Leonardo R. and Tootle, Samuel D. and Hamilton, Maria C. Babiuc and Etienne, Zachariah B.",
    title = "{General relativistic hydrodynamics code for dynamical spacetimes with curvilinear coordinates, tabulated equations of state, and neutrino physics}",
    eprint = "2412.03659",
    archivePrefix = "arXiv",
    primaryClass = "gr-qc",
    doi = "10.1103/hc9l-1thx",
    journal = "Phys. Rev. D",
    volume = "112",
    number = "8",
    pages = "084044",
    year = "2025"
}

@article{Komatsu89b,
	Adsurl = {http://adsabs.harvard.edu/cgi-bin/nph-bib_query?bibcode=1989MNRAS.239..153K&db_key=AST},
	Author = {{Komatsu}, H. and {Eriguchi}, Y. and {Hachisu}, I.},
	Journal = {Mon. Not. R. Astron. Soc.},
	Month = jul,
	Pages = {153-171},
	Title = {{Rapidly rotating general relativistic stars. II -- Differentially rotating polytropes}},
	Volume = 239,
	Year = 1989}

@inproceedings{Kreiss1973MethodsFT,
  title={Methods for the approximate solution of time dependent problems},
  author={Heinz-Otto Kreiss},
  year={1973},
  url={https://api.semanticscholar.org/CorpusID:118627871}
}

@article{Kuan2023a,
    author = "Kuan, Hao-Jui and Lam, Alan Tsz-Lok and Doneva, Daniela D. and Yazadjiev, Stoytcho S. and Shibata, Masaru and Kiuchi, Kenta",
    title = "{Dynamical scalarization during neutron star mergers in scalar-Gauss-Bonnet theory}",
    eprint = "2302.11596",
    archivePrefix = "arXiv",
    primaryClass = "gr-qc",
    doi = "10.1103/PhysRevD.108.063033",
    journal = "Phys. Rev. D",
    volume = "108",
    number = "6",
    pages = "063033",
    year = "2023"
}

@article{Kuan2024,
    author = "Kuan, Hao-Jui and Kiuchi, Kenta and Shibata, Masaru",
    title = "{Tidal Resonance in Binary Neutron Star Inspirals: A High-Precision Study in Numerical Relativity}",
    eprint = "2411.16850",
    archivePrefix = "arXiv",
    primaryClass = "hep-ph",
    doi = "10.1103/j3zk-z17h",
    journal = "Phys. Rev. Lett.",
    volume = "135",
    number = "14",
    pages = "141403",
    year = "2025"
}

@article{Lasota1996,
	Adsurl = {http://adsabs.harvard.edu/abs/1996ApJ...456..300L},
	Author = {{Lasota}, J.-P. and {Haensel}, P. and {Abramowicz}, M.~A.},
	Doi = {10.1086/176650},
	Eprint = {astro-ph/9508118},
	Journal = {Astrophys. J.},
	Month = jan,
	Pages = {300},
	Title = {{Fast Rotation of Neutron Stars}},
	Volume = 456,
	Year = 1996}

@article{Lichnerowicz44,
  author =        {Andre Lichnerowicz},
  journal =       {J. Math. Pures et Appl.},
  key =           {Lichnerowicz44},
  pages =         {37},
  title =         {L'int{\'e}gration des {\'e}quations de la gravitation
                   relativiste et la probl{\`e}me des n corps},
  volume =        {23},
  year =          {1944},
}

@article{LIGOScientific2017vwq,
    author = "Abbott, B. P. and others",
    collaboration = "LIGO Scientific, Virgo",
    title = "{GW170817: Observation of Gravitational Waves from a Binary Neutron Star Inspiral}",
    eprint = "1710.05832",
    archivePrefix = "arXiv",
    primaryClass = "gr-qc",
    reportNumber = "LIGO-P170817",
    doi = "10.1103/PhysRevLett.119.161101",
    journal = "Phys. Rev. Lett.",
    volume = "119",
    number = "16",
    pages = "161101",
    year = "2017"
}

@article{LIGOScientific2018cki,
    author = "Abbott, B. P. and others",
    collaboration = "LIGO Scientific, Virgo",
    title = "{GW170817: Measurements of neutron star radii and equation of state}",
    eprint = "1805.11581",
    archivePrefix = "arXiv",
    primaryClass = "gr-qc",
    reportNumber = "LIGO-P1800115",
    doi = "10.1103/PhysRevLett.121.161101",
    journal = "Phys. Rev. Lett.",
    volume = "121",
    number = "16",
    pages = "161101",
    year = "2018"
}

@article{LIGOScientific2018hze,
    author = "Abbott, B. P. and others",
    collaboration = "LIGO Scientific, Virgo",
    title = "{Properties of the binary neutron star merger GW170817}",
    eprint = "1805.11579",
    archivePrefix = "arXiv",
    primaryClass = "gr-qc",
    doi = "10.1103/PhysRevX.9.011001",
    journal = "Phys. Rev. X",
    volume = "9",
    number = "1",
    pages = "011001",
    year = "2019"
}

@article{Loffler2014,
    author = {L\"offler, Frank and De Pietri, Roberto and Feo, Alessandra and Maione, Francesco and Franci, Luca},
    title = "{Stiffness effects on the dynamics of the bar-mode instability of Neutron Stars in full General Relativity}",
    eprint = "1411.1963",
    archivePrefix = "arXiv",
    primaryClass = "gr-qc",
    doi = "10.1103/PhysRevD.91.064057",
    journal = "Phys. Rev. D",
    volume = "91",
    pages = "064057",
    year = "2015"
}

@ARTICLE{Manca2007,
       author = {{Manca}, Gian Mario and {Baiotti}, Luca and {DePietri}, Roberto and {Rezzolla}, Luciano},
        title = "{Dynamical non-axisymmetric instabilities in rotating relativistic stars}",
      journal = {Classical and Quantum Gravity},
         year = 2007,
        month = jun,
       volume = {24},
       number = {12},
        pages = {S171-S186},
          doi = {10.1088/0264-9381/24/12/S12},
archivePrefix = {arXiv},
       eprint = {0705.1826},
 primaryClass = {astro-ph},
       adsurl = {https://ui.adsabs.harvard.edu/abs/2007CQGra..24S.171M}
}

@article{Markin2023a,
    author = "Markin, Ivan and Neuweiler, Anna and Abac, Adrian and Chaurasia, Swami Vivekanandji and Ujevic, Maximiliano and Bulla, Mattia and Dietrich, Tim",
    title = "{General-relativistic hydrodynamics simulation of a neutron star{\textendash}sub-solar-mass black hole merger}",
    eprint = "2304.11642",
    archivePrefix = "arXiv",
    primaryClass = "gr-qc",
    doi = "10.1103/PhysRevD.108.064025",
    journal = "Phys. Rev. D",
    volume = "108",
    number = "6",
    pages = "064025",
    year = "2023"
}

@article{Musolino2023b,
       author = {{Musolino}, Carlo and {Ecker}, Christian and {Rezzolla}, Luciano},
        title = "{On the Maximum Mass and Oblateness of Rotating Neutron Stars with Generic Equations of State}",
      journal = {Astrophys. J.},
         year = 2024,
        month = feb,
       volume = {962},
       number = {1},
          eid = {61},
        pages = {61},
          doi = {10.3847/1538-4357/ad1758},
archivePrefix = {arXiv},
       eprint = {2307.03225},
 primaryClass = {gr-qc},
       adsurl = {https://ui.adsabs.harvard.edu/abs/2024ApJ...962...61M}
}

@article{Nakamura1987,
    author = "Nakamura, T. and Oohara, K. and Kojima, Y.",
    title = "{General Relativistic Collapse to Black Holes and Gravitational Waves from Black Holes}",
    doi = "10.1143/PTPS.90.1",
    journal = "Prog. Theor. Phys. Suppl.",
    volume = "90",
    pages = "1--218",
    year = "1987"
}

@ARTICLE{Nishad2024,
       author = {{Muhammed}, Nishad and {Duez}, Matthew D. and {Chawhan}, Pavan and {Ghadiri}, Noora and {Buchman}, Luisa T. and {Foucart}, Francois and {Chi-Kit Cheong}, Patrick and {Kidder}, Lawrence E. and {Pfeiffer}, Harald P. and {Scheel}, Mark A.},
        title = "{Stability of hypermassive neutron stars with realistic rotation and entropy profiles}",
      journal = {arXiv e-prints},
         year = 2024,
        month = mar,
          eid = {arXiv:2403.05642},
        pages = {arXiv:2403.05642},
          doi = {10.48550/arXiv.2403.05642},
archivePrefix = {arXiv},
       eprint = {2403.05642},
 primaryClass = {gr-qc},
       adsurl = {https://ui.adsabs.harvard.edu/abs/2024arXiv240305642M}
}

@article{Novak2003,
    author = "Novak, Jerome and Marcq, Emmanuel",
    title = "{Gyromagnetic ratio of rapidly rotating compact stars in general relativity}",
    eprint = "gr-qc/0306010",
    archivePrefix = "arXiv",
    doi = "10.1088/0264-9381/20/14/307",
    journal = "Class. Quant. Grav.",
    volume = "20",
    pages = "3051--3060",
    year = "2003"
}

@article{Ott06c,
	Author = {{Ott}, C.~D. and {Burrows}, A. and {Thompson}, T.~A. and {Livne}, E. and {Walder}, R.},
	Doi = {10.1086/500832},
	Eprint = {astro-ph/0508462},
	Journal = {Astrophys. J. Suppl. Ser.},
	Month = may,
	Pages = {130},
	Title = {{The Spin Periods and Rotational Profiles of Neutron Stars at Birth}},
	Volume = 164,
	Year = 2006}

@article{Palenzuela2025,
    author = "Palenzuela, Carlos and others",
    title = "{MHDuet : a high-order General Relativistic Radiation MHD code for CPU and GPU architectures}",
    eprint = "2510.13965",
    archivePrefix = "arXiv",
    primaryClass = "gr-qc",
    month = "10",
    year = "2025",
    journal = "arXiv"
}

@article{Paschalidis2016,
    author = "Paschalidis, Vasileios and Stergioulas, Nikolaos",
    title = "{Rotating Stars in Relativity}",
    eprint = "1612.03050",
    archivePrefix = "arXiv",
    primaryClass = "astro-ph.HE",
    doi = "10.1007/s41114-017-0008-x",
    journal = "Living Rev. Rel.",
    volume = "20",
    number = "1",
    pages = "7",
    year = "2017"
}

@article{Passamonti2020,
    author = {Passamonti, A and Andersson, N},
    title = "{Merger-inspired rotation laws and the low-T/W instability in neutron stars}",
    journal = {Monthly Notices of the Royal Astronomical Society},
    volume = {498},
    number = {4},
    pages = {5904-5915},
    year = {2020},
    month = {09},
    issn = {0035-8711},
    doi = {10.1093/mnras/staa2725},
    url = {https://doi.org/10.1093/mnras/staa2725},
    eprint = {https://academic.oup.com/mnras/article-pdf/498/4/5904/33838626/staa2725.pdf}
}

@article{Papenfort2021b,
  author =        {Papenfort, L. Jens and Tootle, Samuel D. and
                   Grandcl\'ement, Philippe and Most, Elias R. and
                   Rezzolla, Luciano},
  journal =       {Phys. Rev. D},
  month =         {Jul},
  pages =         {024057},
  publisher =     {American Physical Society},
  title =         {New public code for initial data of unequal-mass,
                   spinning compact-object binaries},
  volume =        {104},
  year =          {2021},
  doi =           {10.1103/PhysRevD.104.024057},
  url =           {https://link.aps.org/doi/10.1103/PhysRevD.104.024057},
}

@article{Pfeiffer:2005,
  author =        {{Pfeiffer}, Harald P. and {York}, James W.},
  journal =       {Phys. Rev. Lett.},
  month =         aug,
  number =        {9},
  pages =         {091101},
  title =         {{Uniqueness and Nonuniqueness in the Einstein
                   Constraints}},
  volume =        {95},
  year =          {2005},
  doi =           {10.1103/PhysRevLett.95.091101},
  eid =           {091101},
}

@article{Pili2017,
    author = {Pili, A. G. and Bucciantini, N. and Drago, A. and Pagliara, G. and Del Zanna, L.},
    title = {Quark deconfinement in the proto-magnetar model of long gamma-ray bursts},
    journal = {Monthly Notices of the Royal Astronomical Society: Letters},
    volume = {462},
    number = {1},
    pages = {L26-L30},
    year = {2016},
    month = {06},
    issn = {1745-3925},
    doi = {10.1093/mnrasl/slw115},
    url = {https://doi.org/10.1093/mnrasl/slw115}
}

@article{Rainho:2025ykl,
    author = "Rainho, In{\^e}s and Bamber, Jamie and Guerra, Davide and Miravet-Ten{\'e}s, Miquel and Ruiz, Milton and Tsokaros, Antonios and Shapiro, Stuart L.",
    title = "{Impact of Magnetic Field Topology on Electromagnetic and Gravitational Waves from Binary Neutron Star Merger Remnants}",
    eprint = "2510.17511",
    archivePrefix = "arXiv",
    primaryClass = "astro-ph.HE",
    month = "10",
    year = "2025",
    journal = "arXiv"
}

@article{Rashti2021,
  author =        {Rashti, Alireza and Fabbri, Francesco Maria and
                   Br\"ugmann, Bernd and Chaurasia, Swami Vivekanandji and
                   Dietrich, Tim and Ujevic, Maximiliano and
                   Tichy, Wolfgang},
  journal =       {Phys. Rev. D},
  number =        {10},
  pages =         {104027},
  title =         {{New pseudospectral code for the construction of
                   initial data}},
  volume =        {105},
  year =          {2022},
  doi =           {10.1103/PhysRevD.105.104027},
}

@article{Rashti2024,
    author = "Rashti, Alireza and Noe, Andrew",
    title = "{Realistic binary neutron star initial data with Elliptica}",
    eprint = "2407.01701",
    archivePrefix = "arXiv",
    primaryClass = "gr-qc",
    doi = "10.1088/1361-6382/ad942d",
    journal = "Class. Quant. Grav.",
    volume = "42",
    number = "1",
    pages = "015007",
    year = "2025"
}

@article{Reboul-Salze:2024jst,
    author = {Reboul-Salze, Alexis and Barr{\`e}re, Paul and Kiuchi, Kenta and Guilet, J{\'e}r{\^o}me and Raynaud, Rapha{\"e}l and Fujibayashi, Sho and Shibata, Masaru},
    title = "{Tayler-Spruit dynamo in binary neutron star merger remnants}",
    eprint = "2411.19328",
    archivePrefix = "arXiv",
    primaryClass = "astro-ph.HE",
    doi = "10.1051/0004-6361/202453126",
    journal = "Astron. Astrophys.",
    volume = "699",
    pages = "A4",
    year = "2025"
}

@article{Rezzolla00,
	Adsurl = {http://adsabs.harvard.edu/abs/2000ApJ...531L.139R},
	Author = {{Rezzolla}, L. and {Lamb}, F.~K. and {Shapiro}, S.~L.},
	Doi = {10.1086/312539},
	Eprint = {astro-ph/9911188},
	Journal = {Astrophys. J. Lett.},
	Month = mar,
	Pages = {L139-L142},
	Title = {{R-Mode Oscillations in Rotating Magnetic Neutron Stars}},
	Volume = 531,
	Year = 2000}

@inbook{Rosswog2024a,
    author = "Rosswog, Stephan and Diener, Peter",
    title = "{SPHINCS{\_}BSSN: Numerical Relativity with~Particles}",
    eprint = "2404.15952",
    archivePrefix = "arXiv",
    primaryClass = "gr-qc",
    doi = "10.1007/978-981-97-8522-3_7",
    year = "2025"
}

@article{Saijo2003,
	Adsurl = {http://adsabs.harvard.edu/abs/2003ApJ...595..352S},
	Author = {{Saijo}, M. and {Baumgarte}, T.~W. and {Shapiro}, S.~L.},
	Doi = {10.1086/377334},
	Eprint = {astro-ph/0302436},
	Journal = {Astrophys. J.},
	Month = sep,
	Pages = {352-364},
	Title = {{One-armed Spiral Instability in Differentially Rotating Stars}},
	Volume = 595,
	Year = 2003}

@article{Shibata1995,
    author = "Shibata, Masaru and Nakamura, Takashi",
    title = "{Evolution of three-dimensional gravitational waves: Harmonic slicing case}",
    doi = "10.1103/PhysRevD.52.5428",
    journal = "Phys. Rev. D",
    volume = "52",
    pages = "5428--5444",
    year = "1995"
}

@article{Shibata2000,
    author = "Shibata, Masaru and Baumgarte, Thomas W. and Shapiro, Stuart L.",
    title = "{The bar-mode instability in differentially rotating neutron stars: simulations in full general relativity}",
    eprint = "astro-ph/0005378",
    archivePrefix = "arXiv",
    doi = "10.1086/309525",
    journal = "Astrophys. J.",
    volume = "542",
    pages = "453--463",
    year = "2000"
}

@article{Shibata:2002mr,
	Author = {Shibata, Masaru and Karino, Shigeyuki and Eriguchi, Yoshiharu},
	Eprint = {gr-qc/0206002},
	Journal = {Mon. Not. R. Astron. Soc.},
	Pages = {L27},
	Slaccitation = {%%CITATION = GR-QC 0206002;%%},
	Title = {Dynamical instability of differentially rotating stars},
	Volume = {334},
	Year = {2002}}

@article{Shibata2003ga,
	Adsurl = {http://adsabs.harvard.edu/abs/2003PhRvD..68h4020S},
	Author = {{Shibata}, M. and {Taniguchi}, K. and {Ury{\=u}}, K.},
	Doi = {10.1103/PhysRevD.68.084020},
	Eid = {084020},
	Eprint = {gr-qc/0310030},
	Journal = {Phys. Rev. D},
	Month = oct,
	Number = 8,
	Pages = {084020},
	Title = {{Merger of binary neutron stars of unequal mass in full general relativity}},
	Volume = 68,
	Year = 2003}

@article{Schnetter2003,
    author = "Schnetter, Erik and Hawley, Scott H. and Hawke, Ian",
    title = "{Evolutions in 3-D numerical relativity using fixed mesh refinement}",
    eprint = "gr-qc/0310042",
    archivePrefix = "arXiv",
    reportNumber = "AEI-2003-078",
    doi = "10.1088/0264-9381/21/6/014",
    journal = "Class. Quant. Grav.",
    volume = "21",
    pages = "1465--1488",
    year = "2004"
}

@article{Staykov2023,
       author = {{Staykov}, Kalin V. and {Doneva}, Daniela D. and {Heisenberg}, Lavinia and {Stergioulas}, Nikolaos and {Yazadjiev}, Stoytcho S.},
        title = "{Differentially rotating scalarized neutron stars with realistic postmerger profiles}",
      journal = {Phys. Rev. D},
         year = 2023,
        month = jul,
       volume = {108},
       number = {2},
          eid = {024058},
        pages = {024058},
          doi = {10.1103/PhysRevD.108.024058},
archivePrefix = {arXiv},
       eprint = {2303.07769},
 primaryClass = {gr-qc},
       adsurl = {https://ui.adsabs.harvard.edu/abs/2023PhRvD.108b4058S}
}

@article{Stergioulas95,
	Adsurl = {http://adsabs.harvard.edu/abs/1995ApJ...444..306S},
	Author = {{Stergioulas}, N. and {Friedman}, J.~L.},
	Doi = {10.1086/175605},
	Eprint = {astro-ph/9411032},
	Journal = {Astrophys. J.},
	Month = may,
	Pages = {306-311},
	Title = {{Comparing models of rapidly rotating relativistic stars constructed by two numerical methods}},
	Volume = 444,
	Year = 1995}

@article{Stergioulas04,
	Adsurl = {http://adsabs.harvard.edu/abs/2004MNRAS.352.1089S},
	Author = {{Stergioulas}, N. and {Apostolatos}, T.~A. and {Font}, J.~A},
	Doi = {10.1111/j.1365-2966.2004.07973.x},
	Eprint = {arXiv:astro-ph/0312648},
	Journal = {Mon. Not. R. Astron. Soc.},
	Month = aug,
	Pages = {1089-1101},
	Title = {{Non-linear pulsations in differentially rotating neutron stars: mass-shedding-induced damping and splitting of the fundamental mode}},
	Volume = 352,
	Year = 2004}

@article{Studzinska2016,
	Adsurl = {http://adsabs.harvard.edu/abs/2016MNRAS.463.2667S},
	Author = {{Studzi{\'n}ska}, A.~M. and {Kucaba}, M. and {Gondek-Rosi{\'n}ska}, D. and {Villain}, L. and {Ansorg}, M.},
	Doi = {10.1093/mnras/stw2152},
	Journal = {Mon. Not. R. Astron. Soc.},
	Month = dec,
	Pages = {2667-2679},
	Title = {{Effect of the equation of state on the maximum mass of differentially rotating neutron stars}},
	Volume = 463,
	Year = 2016}

@article{Takami2014,
	Adsurl = {http://adsabs.harvard.edu/abs/2014PhRvL.113i1104T},
	Archiveprefix = {arXiv},
	Author = {{Takami}, K. and {Rezzolla}, L. and {Baiotti}, L.},
	Doi = {10.1103/PhysRevLett.113.091104},
	Eid = {091104},
	Eprint = {1403.5672},
	Journal = {Phys. Rev. Lett.},
	Month = aug,
	Number = 9,
	Pages = {091104},
	Primaryclass = {gr-qc},
	Title = {{Constraining the Equation of State of Neutron Stars from Binary Mergers}},
	Volume = 113,
	Year = 2014}

@ARTICLE{Uryu2017,
       author = {{Ury{\={u}}}, K{\={o}}ji and {Tsokaros}, Antonios and {Baiotti}, Luca and {Galeazzi}, Filippo and {Taniguchi}, Keisuke and {Yoshida}, Shin'ichirou},
        title = "{Modeling differential rotations of compact stars in equilibriums}",
      journal = {Phys. Rev. D},
         year = 2017,
        month = nov,
       volume = {96},
       number = {10},
          eid = {103011},
        pages = {103011},
          doi = {10.1103/PhysRevD.96.103011},
archivePrefix = {arXiv},
       eprint = {1709.02643},
 primaryClass = {astro-ph.HE},
       adsurl = {https://ui.adsabs.harvard.edu/abs/2017PhRvD..96j3011U}
}

@article{Uryu2019,
    author = "Uryu, Koji and Yoshida, Shijun and Gourgoulhon, Eric and Markakis, Charalampos and Fujisawa, Kotaro and Tsokaros, Antonios and Taniguchi, Keisuke and Eriguchi, Yoshiharu",
    title = "{New code for equilibriums and quasiequilibrium initial data of compact objects. IV. Rotating relativistic stars with mixed poloidal and toroidal magnetic fields}",
    eprint = "1906.10393",
    archivePrefix = "arXiv",
    primaryClass = "gr-qc",
    doi = "10.1103/PhysRevD.100.123019",
    journal = "Phys. Rev. D",
    volume = "100",
    number = "12",
    pages = "123019",
    year = "2019"
}

@ARTICLE{Weih2017,
   author = {{Weih}, L.~R. and {Most}, E.~R. and {Rezzolla}, L.},
    title = "{On the stability and maximum mass of differentially rotating relativistic stars}",
  journal = {Mon. Not. R. Astron. Soc.},
archivePrefix = "arXiv",
   eprint = {1709.06058},
 primaryClass = "gr-qc",
     year = 2018,
    month = jan,
   volume = 473,
    pages = {L126-L130},
      doi = {10.1093/mnrasl/slx178},
   adsurl = {http://adsabs.harvard.edu/abs/2018MNRAS.473L.126W}
}

@article{Werneck2022,
    author = "Werneck, Leonardo R. and others",
    title = "{Addition of tabulated equation of state and neutrino leakage support to illinoisgrmhd}",
    eprint = "2208.14487",
    archivePrefix = "arXiv",
    primaryClass = "gr-qc",
    doi = "10.1103/PhysRevD.107.044037",
    journal = "Phys. Rev. D",
    volume = "107",
    number = "4",
    pages = "044037",
    year = "2023"
}

@article{Xie2020,
  title = {Instabilities in neutron-star postmerger remnants},
  author = {Xie, Xiaoyi and Hawke, Ian and Passamonti, Andrea and Andersson, Nils},
  journal = {Phys. Rev. D},
  volume = {102},
  issue = {4},
  pages = {044040},
  numpages = {17},
  year = {2020},
  month = {Aug},
  publisher = {American Physical Society},
  doi = {10.1103/PhysRevD.102.044040},
  url = {https://link.aps.org/doi/10.1103/PhysRevD.102.044040}
}

@article{York73,
  author =        {York, James W.},
  journal =       {J. Math. Phys.},
  pages =         {456},
  title =         {Conformally invariant orthogonal decomposition of
                   symmetric tensors on Riemannian manifolds and the
                   initial value problem of general relativity},
  volume =        {14},
  year =          {1973},
}

@article{York99,
  author =        {James W. York},
  journal =       {Phys. Rev. Lett.},
  pages =         {1350-1353},
  title =         {Conformal `thin-sandwich' data for the initial-value
                   problem of general relativity},
  volume =        {82},
  year =          {1999},
}

@article{Zhou2019,
  title = {Differentially rotating strange star in general relativity},
  author = {Zhou, Enping and Tsokaros, Antonios and Uryu, Koji and Xu, Renxin and Shibata, Masaru},
  journal = {Phys. Rev. D},
  volume = {100},
  issue = {4},
  pages = {043015},
  numpages = {11},
  year = {2019},
  month = {Aug},
  publisher = {American Physical Society},
  doi = {10.1103/PhysRevD.100.043015},
  url = {https://link.aps.org/doi/10.1103/PhysRevD.100.043015}
}

@misc{GRHayL,
  key =           {GRHayL},
  note =          {General Relativistic Hydrodynamic Library,
                   \url{https://github.com/GRHayL/GRHayL}},
  title =         {General Relativistic Hydrodynamic Library},
  url =           {https://github.com/GRHayL/GRHayL},
}

@misc{lorene,
  key =           {LORENE},
  note =          {Langage Objet pour la RElativit\'{e} Num\'{e}rique,
                   \url{www.lorene.obspm.fr}},
  title =         {Langage Objet pour la RElativit\'{e} Num\'{e}rique},
  url =           {http://www.lorene.obspm.fr},
}

@misc{fukascience,
  key =           {fukascience},
  title =         {FUKA Enabled Science},
  url =           {https://samueltootle.github.io/fuka/fukascience.html},
}

@misc{stellarcollapse,
  key =           {stellarcollapse},
  title =         {Stellar Collapse},
  url =           {https://stellarcollapse.org/},
}

@Misc{EinsteinToolkit:2025_05,
  requested-for ={EinsteinToolkit},
  author       = {Maxwell Rizzo and Roland Haas and Steven R. Brandt and Zachariah Etienne and Deborah Ferguson and Lucas Timotheo Sanches and Bing-Jyun Tsao and Leonardo Werneck and David Boyer and Gabriele Bozzola and Cheng-Hsin Cheng and Samuel Cupp and Peter Diener and Terrence Pierre Jacques and Liwei Ji and Hayley Macpherson and Ivan Markin and Erik Schnetter and Wolfgang Tichy and Samuel Tootle and Yumeng Xu and Miguel Zilhão and Yosef Zlochower and Miguel Alcubierre and Daniela Alic and Gabrielle Allen and Marcus Ansorg and Federico G. Lopez Armengol and Maria Babiuc-Hamilton and Luca Baiotti and Werner Benger and Eloisa Bentivegna and Sebastiano Bernuzzi and Krishiv Bhatia and Tanja Bode and Brockton Brendal and Bernd Bruegmann and Manuela Campanelli and Michail Chabanov and Federico Cipolletta and Giovanni Corvino and Roberto De Pietri and Alexandru Dima and Harry Dimmelmeier and Jake Doherty and Rion Dooley and Nils Dorband and Matthew Elley and Yaakoub El Khamra and Lorenzo Ennoggi and Joshua Faber and Giuseppe Ficarra and Toni Font and Joachim Frieben and Bruno Giacomazzo and Tom Goodale and Carsten Gundlach and Ian Hawke and Scott Hawley and Ian Hinder and E. A. Huerta and Sascha Husa and Taishi Ikeda and Sai Iyer and Daniel Johnson and Abhishek V. Joshi and Jay Kalinani and Anuj Kankani and Wolfgang Kastaun and Thorsten Kellermann and Andrew Knapp and Michael Koppitz and Pablo Laguna and Gerd Lanferman and Paul Lasky and Frank Löffler and Joan Masso and Lars Menger and Andre Merzky and Jonah Maxwell Miller and Mark Miller and Philipp Moesta and Pedro Montero and Bruno Mundim and Patrick Nelson and Andrea Nerozzi and Scott C. Noble and Christian Ott and Ludwig Jens Papenfort and Ravi Paruchuri and Michal Pirog and Denis Pollney and Daniel Price and David Radice and Thomas Radke and Christian Reisswig and Luciano Rezzolla and Chloe B. Richards and David Rideout and Matei Ripeanu and Lorenzo Sala and Jascha A Schewtschenko and Bernard Schutz and Ed Seidel and Eric Seidel and John Shalf and Swapnil Shankar and Ken Sible and Ulrich Sperhake and Nikolaos Stergioulas and Wai-Mo Suen and Bela Szilagyi and Ryoji Takahashi and Michael Thomas and Jonathan Thornburg and Chi Tian and Malcolm Tobias and Aaryn Tonita and Paul Walker and Mew-Bing Wan and Barry Wardell and Helvi Witek and Burkhard Zink},
  title        = {The {E}instein {T}oolkit},
  month        = may,
  year         = 2025,
  publisher    = {Zenodo},
  version      = {The "Martin D. Kruskal" release, ET\_2025\_05},
  doi          = {10.5281/zenodo.15520463},
  url          = {https://doi.org/10.5281/zenodo.15520463},
}

@Article{NumPy,
  title         = {Array programming with {NumPy}},
  author        = {Charles R. Harris and K. Jarrod Millman and St{\'{e}}fan J.
              van der Walt and Ralf Gommers and Pauli Virtanen and David
              Cournapeau and Eric Wieser and Julian Taylor and Sebastian
              Berg and Nathaniel J. Smith and Robert Kern and Matti Picus
              and Stephan Hoyer and Marten H. van Kerkwijk and Matthew
              Brett and Allan Haldane and Jaime Fern{\'{a}}ndez del
              R{\'{i}}o and Mark Wiebe and Pearu Peterson and Pierre
              G{\'{e}}rard-Marchant and Kevin Sheppard and Tyler Reddy and
              Warren Weckesser and Hameer Abbasi and Christoph Gohlke and
              Travis E. Oliphant},
  year          = {2020},
  month         = sep,
  journal       = {Nature},
  volume        = {585},
  number        = {7825},
  pages         = {357--362},
  doi           = {10.1038/s41586-020-2649-2},
  publisher     = {Springer Science and Business Media {LLC}},
}

@Article{Matplotlib,
  Author    = {Hunter, J. D.},
  Title     = {Matplotlib: A 2D graphics environment},
  Journal   = {Computing in Science \& Engineering},
  Volume    = {9},
  Number    = {3},
  Pages     = {90--95},
  publisher = {IEEE COMPUTER SOC},
  doi       = {10.1109/MCSE.2007.55},
  year      = 2007
}

@article{Kuibit,
    author = "Bozzola, Gabriele",
    title = "{kuibit: Analyzing Einstein Toolkit simulations with Python}",
    eprint = "2104.06376",
    archivePrefix = "arXiv",
    primaryClass = "gr-qc",
    doi = "10.21105/joss.03099",
    journal = "J. Open Source Softw.",
    volume = "6",
    number = "60",
    pages = "3099",
    year = "2021"
}

@software{spectrecode,
    author = "Deppe, Nils and Throwe, William and Kidder, Lawrence E. and Vu,
Nils L. and Nelli, Kyle C. and Armaza, Crist\'obal and Bonilla, Marceline S. and
H\'ebert, Fran\c{c}ois and Kim, Yoonsoo and Kumar, Prayush and Lovelace,
Geoffrey and Macedo, Alexandra and Moxon, Jordan and O'Shea, Eamonn and
Pfeiffer, Harald P. and Scheel, Mark A. and Teukolsky, Saul A. and Wittek,
Nikolas A. and others",
    title = "\texttt{SpECTRE v2025.08.19}",
    version = "2025.08.19",
    publisher = "Zenodo",
    doi = "10.5281/zenodo.16906840",
    url = "https://spectre-code.org",
    howpublished =
"\href{https://doi.org/10.5281/zenodo.16906840}{10.5281/zenodo.16906840}",
    license = "MIT",
    year = "2025",
    month = "8"
}

\end{document}